\documentclass[twocolumn,aps]{revtex4-1}

\usepackage{amsmath}
\usepackage{bm}
\usepackage{color, xcolor}
\usepackage{graphicx}
\usepackage{hyperref}

\newcommand{\B}[1]{{\bm{#1}}}

\usepackage[latin1]{inputenc}
\newcommand{\beq}{\begin{equation}}
\newcommand{\eeq}{\end{equation}}
\newcommand{\bea}{\begin{eqnarray}}
\newcommand{\eea}{\end{eqnarray}}
\newcommand{\pa}{\partial}

\renewcommand{\=}{\!=\!}
\newcommand{\avg}[1]{\left\langle #1 \right\rangle}
\def\bigstrut{\rule[-1ex]{0pt}{3.5ex}}

\hypersetup{
   colorlinks,
   linkcolor={blue!100!black},
   citecolor={green!50!black},
   urlcolor={blue!80!black}
}

\begin{document}

\title{Necking instabilities in elasto-viscoplastic materials}
\author{Avraham Moriel and Eran Bouchbinder}
\affiliation{Chemical and Biological Physics Department, Weizmann Institute of Science, Rehovot 7610001, Israel}


\begin{abstract}
Necking instabilities, in which tensile (extensional) deformation localizes into a small spatial region, are generic failure modes in elasto-viscoplastic materials. Materials in this very broad class --- including amorphous, crystalline, polycrystalline and other materials --- feature a predominantly elastic response at small stresses, plasticity onset at a rather well-defined yield stress and rate-dependence. Necking instabilities involve a unique coupling between the system's geometry and its constitutive behavior. We consider generic elasto-viscoplastic constitutive relations involving an internal-state field, which represents the structural evolution of the material during plastic deformation, and study necking in the long-wavelength approximation (sometimes termed the lubrication or slender-bar approximation). We derive a general expression for the largest time-dependent eigenvalue in an approximate WKB-like linear stability analysis, highlighting various basic physical effects involved in necking. This expression is then used to propose criteria for the onset of necking and more importantly for the emergence of strong localization. Applications to strain-softening amorphous plasticity, in the framework of the Shear-Transformation-Zone model, and to strain-hardening crystalline/polycrystalline plasticity, in the framework of the Kocks-Mecking model, are presented. These quantitative analyses of widely different material models support the theoretical predictions, most notably for the strong localization during necking.
\end{abstract}
\maketitle

\section{Introduction}
\label{sec:intro}

\begin{figure}[h!]
\includegraphics[width=0.48\textwidth]{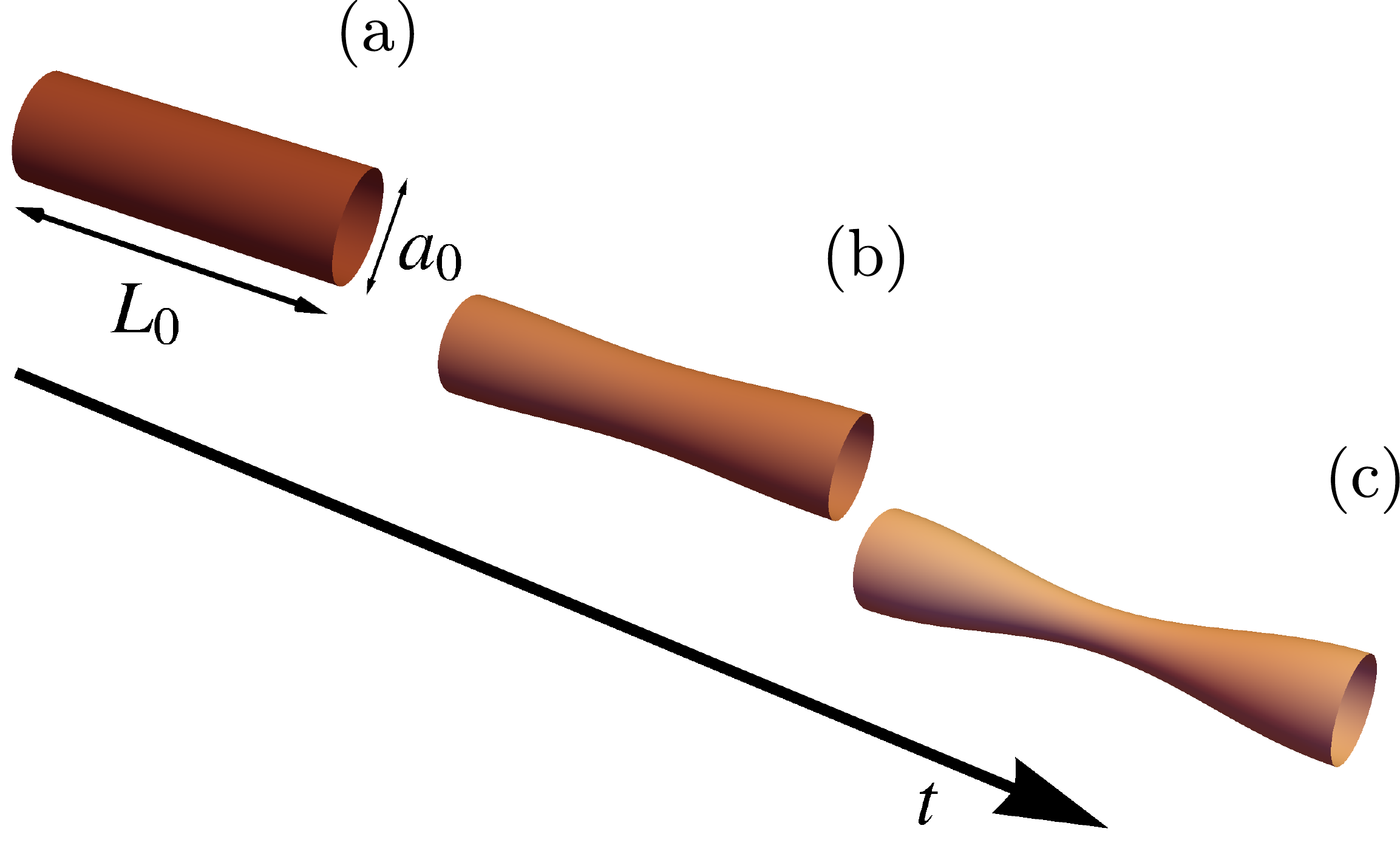}
\caption{A schematic illustration of the development of a necking instability of an initially spatially-homogeneous cylindrical bar of length $L_0$ and cross-sectional area $a_0$ (panel a). Extensional deformation leads to the onset of a neck, i.e.~region in which the cross-sectional area is slightly reduced compared to the rest of the bar (panel b). As the extensional deformation proceeds, the necking instability further develops, and the bar experiences strong localization (panel c) that is likely to lead to catastrophic failure.}
\label{fig:figure1}
\end{figure}

The ability of materials and structures to withstand tensile forces without failure is a fundamental physical property with far-reaching practical implications. In elasto-viscoplastic materials --- a very broad class of materials which feature a predominantly elastic response at small stresses, irreversible plastic deformation upon surpassing a rather well-defined yield stress and rate-dependence --- the major process that limits this ability is the development of necking instabilities, observed in a broad range of materials such as glassy alloys (bulk metallic glasses)~\cite{Guo2007,Hofmann2008,Vormelker2008}, many crystalline/polycrystalline materials~\cite{Bridgman1953,Bridgman1964}, emulsions and suspensions~\cite{Clasen2003,Niedzwiedz2010,Szabo2012,Huisman2012,Louvet2014}, and amorphous bubble rafts~\cite{Kuo2012,Kuo2013,Arciniaga2011}. This generic instability manifests itself in the form of strongly localized deformation that results in a significant local reduction in the material's cross-sectional area, which typically leads to failure, cf.~Fig.~\ref{fig:figure1}. Consequently, understanding and predicting necking instabilities are of great importance as is also reflected by the quite extensive literature devoted to the topic, dating back at least to Consid\`ere's 1885 work~\cite{Considere1885}.

In addition to the practical importance of necking instabilities in elasto-viscoplastic materials, this problem also poses basic scientific challenges. One essential element of the problem is the viscoplastic constitutive relation of the material, which generically involves strongly nonlinear and far-from-equilibrium dynamics, coupled to the irreversible structural evolution of the material. Such constitutive relations are not yet fully developed and consequently their implications for necking instabilities are not yet fully understood.

Another essential element of the problem is the macroscopic material geometry (e.g.~a long bar which is initially cylindrical) and its dynamic evolution due to extensional driving forces (e.g.~stretching along the cylinder's major axis). In particular, tensile/extensional deformation --- whether spatially-homogeneous or inhomogeneous --- generically leads to a reduction in the material cross-sectional area, which in turn leads to enhanced stresses that may drive additional tensile/extensional deformation. This intrinsic coupling between geometry and the constitutive response of the material gives rise to rich physical behaviors that call for theoretical understanding.

The macroscopic geometry of the material and its dynamic evolution have other important implications which make the problem interesting and challenging. Most notably, they imply that spatially-homogeneous deformation must be intrinsically time-dependent, since even steady extensional deformation is accompanied by a continuous (time-dependent) reduction in the cross-sectional area. This, in turn, implies that a conventional linear stability analysis --- the standard tool for studying instabilities in a broad range of physical problems --- does not strictly apply to this problem; if some approximate form of it does apply and some relevant eigenvalue problem for the growth of shape perturbations can be derived, then the emerging eigenvalues and eigenvectors are time-dependent.

Necking instabilities have been quite extensively studied in various scientific contexts and communities. While we cannot systematically and exhaustively review the relevant literature here, we would like to mention a few works that provide some background to our work. We mainly focus on analytical works, which typically invoke the long-wavelength approximation (sometimes termed the lubrication or slender-bar approximation) in which shape perturbations along the extensional deformation axis are characterized by a wavelength that is much larger than the lateral dimensions of the system. In the context of rate-independent plasticity models, the first criterion for the onset of necking has been proposed by Consid\`ere~\cite{Considere1885}. Various analyses followed, see for example the early review by Orowan~\cite{Orowan1949} and many other later works~\cite{Needleman1972,Hutchinson1978,Zaera2014,Audoly2016,Rubin2017}. Later, Hart extended Consid\`ere's criterion to include material rate-sensitivity~\cite{Hart1967}, and many analyses followed, e.g.~\cite{Campbell1967,Jonas1976,Aragon1973}. Essentially all of these works focussed on the onset of necking using linearized analyses. A few other works~\cite{Marciniak1967,Marciniak1973, Hutchinson1977, Fressengeas1985}, however, considered a particular class of phenomenological plasticity models that enabled the development of a nonlinear stability analysis.

Necking instabilities in extensional flows of non-Newtonian fluids, such as polymer melts, have also been widely studied, see the review papers of~\cite{Ide1976,Denn2001}. Recently, necking instabilities in various models of complex fluids and soft solids have been considered~\cite{Fielding2011,Hoyle2015,Hoyle2016a,Hoyle2016b}. The vast majority of the literature on necking instabilities does not take into account the structural evolution of the material during plastic deformation; a contrary example is the very recent works of~\cite{Yasnikov2014,Yasnikov2017} that considered phenomenological models of crystalline plasticity in which the dislocation density evolves with plastic deformation and affects it.

In this work, we study necking instabilities in elasto-viscoplastic materials in the long-wavelength approximation, within a generic constitutive framework that accounts for the intrinsic rate-dependence of plastic deformation and for the structural evolution of the material during plastic deformation through an internal-state field. We derive a general expression for the largest time-dependent eigenvalue in an approximated WKB-like linear stability analysis, allowing us to identify the various stabilizing and destabilizing physical processes involved. The resulting expression can be used to predict necking instabilities in a broad range of materials and constitutive relations.

Unlike most of the previous works, we do not only consider the onset of instability, but also extensively discuss the emergence of strong localization. A criterion for the latter is obtained by comparing the time-evolution of shape perturbations and the time-evolution of the spatially-homogeneous (unperturbed) state. The criterion is applied to two very different rate-dependent constitutive relations that feature an internal-state field: the Shear-Transformation-Zone (STZ) model~\cite{Falk1997,Bouchbinder2009a,Bouchbinder2009b,Bouchbinder2009c,Falk2010} of strain-softening amorphous plasticity and the Kocks-Mecking model~\cite{Mecking1981,Estrin1984,Follansbee1988,Kocks2003} of strain-hardening crystalline/polycrystalline plasticity. Comparison of the strong localization criterion to nonlinear numerical solutions of the two different models demonstrates favorable agreement, lending support to the proposed criterion.

The structure of this paper is as follows; the complete mathematical formulation of the necking problem and its dimensionally-reduced form are presented in Sect.~\ref{sec:formulation}. The time-dependent linear stability analysis is presented in Sect.~\ref{sec:LSA}, and the derivation of a general analytical expression for the largest time-dependent eigenvalue and the associated onset criterion for necking follow in Sect.~\ref{sec:largest_eigenvalue}. The condition for the emergence of strong localization is derived in Sect.~\ref{sec:localization}. The theoretical predictions are tested against full numerical solutions for amorphous/glassy materials (using the Shear-Transformation-Zone model) in Sect.~\ref{sec:STZ} and for crystalline/polycrystalline materials (using the Kocks-Mecking model) in Sect.~\ref{sec:KM}. Finally, some concluding remarks and future research directions are offered in Sect.~\ref{sec:summary}.

\section{Problem formulation: Conservation laws, constitutive framework and the long-wavelength approximation}
\label{sec:formulation}

We begin by mathematically formulating the problem, i.e.~writing down the relevant conservation laws, defining a generic constitutive framework for elasto-viscoelastic materials with an internal-state field and stating the invoked approximations. Mass conservation is described by the continuity equation
\begin{equation}
\label{eq:mass_conservation}
  \partial_t \rho + \nabla\!\cdot\!\left({\B v} \rho\right)=0 \ ,
\end{equation}
where $\rho({\B r},t)$ is the mass density, i.e.~the mass per unit volume at point $\B r$ in space at time $t$, and ${\B v}({\B r},t)$ is the velocity field (here and below vectors and tensors appear in bold face). Linear momentum conservation takes the form
\begin{equation}
\label{eq:momentum_balance}
  \nabla\!\cdot\!\B\sigma  = \rho\,\dot{\B v} \ ,
\end{equation}
where ${\B \sigma}({\B r},t)$ is the Cauchy (real) stress tensor and body forces were neglected. Note that $\dot{\bullet}\!\equiv\!\left(\partial_t+{\B v}\!\cdot\!\nabla\right)\!\bullet$, is the full (material) time derivative, and that ${\B v}\!\equiv\!\partial_t\B u$, where $\B u(\B r,t)$ is the displacement field. Finally, angular momentum conservation is ensured by the symmetry of the Cauchy stress tensor $\bm{\sigma}\!=\!\bm{\sigma}^{T}$~\cite{Landau7}.

In order to make reference to a specific class of materials and to render the problem mathematically well-defined, we need to specify a relation between the stress ${\B \sigma}$ and the velocity $\B v$, i.e.~we need a constitutive law. More precisely, we need a relation between the total strain-rate (the symmetric part of the velocity gradient tensor) $\dot{\bm{\epsilon}}\!\equiv\!\tfrac{1}{2}[\nabla {\B v}+(\nabla {\B v})^{T}]$ and $\B \sigma$ that is representative of elasto-viscoplastic materials. As viscoplasticity is an intrinsically dynamic, rate-dependent class of physical phenomena, it must be expressed in terms of the strain-rate $\dot{\bm{\epsilon}}$.

We proceed in two steps; first, we decompose the total strain-rate into an elastic (reversible) contribution $\dot{\bm{\epsilon}}^{el}$ and a viscoplastic (irreversible) contribution $\dot{\bm{\epsilon}}^{pl}$,
\begin{equation}
\label{eq:strain_rate_decomp}
  \dot{\bm{\epsilon}}=\dot{\bm{\epsilon}}^{el}+\dot{\bm{\epsilon}}^{pl} \ .
\end{equation}
Second, we discuss the structure of the elastic and viscoplastic contributions. For the elastic part, we use Hooke's law of isotropic linear elasticity $\B\epsilon(\B\sigma)$~\cite{Landau7}, and take a proper time derivative to cast it in rate form~\cite{Xiao2006}. Consequently, $\dot{\bm{\epsilon}}^{el}\!\propto\!\dot{\B\sigma}$ in the tensorial sense. For the viscoplastic part, we leave the functional form of $\dot{\bm{\epsilon}}^{pl}(\cdots)$ unspecified, and focus only on its arguments. It obviously depends on the stress $\B\sigma$ and it must depend on the structural state of the material that is captured by a small set of properly-defined, coarse-grained internal-state fields ${\cal I}_\alpha$. In addition, as viscoplastic deformation might involve thermal activation, it might depend on the temperature $T$.

$\dot{\bm{\epsilon}}^{pl}\left(\bm{\sigma},T,{\cal I}_{\alpha}\right)$ should be supplemented by evolution equations for the internal-state fields ${\cal I}_\alpha$, which are taken here to be scalar for simplicity (though non-scalar internal-state fields are sometimes essential, e.g.~in the context of the Bauschinger effect~\cite{Bauschinger1881}). As no structural evolution takes place in the absence of viscoplastic deformation, we must have
\begin{equation}
\label{eq:internal_variable}
  \dot{\cal I}_\alpha=\dot{\bm{\epsilon}}^{pl}\!\!:\!{\B g}\!\left(\bm{\sigma},{\cal I}_\alpha\right) \ ,
\end{equation}
where ${\B g}(\cdots)$ is a tensorial function that renders the right-hand-side of Eq.~\eqref{eq:internal_variable} a proper scalar. It is natural and physically intuitive to expect $\B g\!\propto\!\B\sigma$ such that ${\cal I}_\alpha$ evolves due to the viscoplastic dissipation rate $\dot{\bm{\epsilon}}^{pl}\!\!:\!\B\sigma$. However, as not all internal-state viscoplastic models available in the literature share this feature, we do not restrict the discussion to this case. Finally, we note that Eq.~\eqref{eq:strain_rate_decomp} $\dot{\bm{\epsilon}}(\B v)\!=\!\dot{\bm{\epsilon}}^{el}(\dot{\B\sigma})+\dot{\bm{\epsilon}}^{pl}\!\left(\bm{\sigma},T,{\cal I}_{\alpha}\right)$ is in fact an equation for $\dot{\B\sigma}$.

Once $\dot{\bm{\epsilon}}^{pl}\!\left(\bm{\sigma},T,{\cal I}_{\alpha}\right)$ and ${\B g}\!\left(\bm{\sigma},{\cal I}_\alpha\right)$ are specified for a given class of materials, Eqs.~\eqref{eq:mass_conservation}-\eqref{eq:internal_variable} fully describe the evolution of $\rho$, $\B v$, $\B\sigma$ and ${\cal I}_\alpha$  for any initial-boundary value problem with any initial geometry (if the temperature $T$ evolves in space and time, then the heat equation should be added). Applying this general framework to the necking problem, i.e.~to the large elasto-viscoplastic extensional deformation of a cylindrical bar, is an extremely complicated mathematical problem, even if only numerical solutions are considered.

In order to gain some analytical and physical insight into this complicated problem we invoke several approximations. First, we note that viscoplastic deformation is quite generically  a slow process compared to elastic processes (i.e.~to the relevant travel time of elastic waves). Consequently, as long as we do not consider high applied strain-rates, we can approximate Eq.~\eqref{eq:momentum_balance} by its quasi-static counterpart $\nabla\cdot\bm{\sigma}\!=\!{\bm 0}$. By so doing, we exclude from the discussion a class of interesting necking problems in which inertial effects play an important role~\cite{Mercier2003,Rittel2010,Osovski2012,Osovski2013,Osovski2015,Zaera2014,Vaz-Romero2016,Vaz-Romero2017a,Rodriguez-Martinez2017a}. We further assume that the material of interest is incompressible, i.e.~its mass density is constant (and hence is omitted hereafter), which is a reasonably good approximation for most elasto-viscoplastic materials.

Next, we consider a long cylinder with an initial length $L_0$ and cross-sectional area $a_0$, cf.~Fig.~\ref{fig:figure1}a, such that its initial radius satisfies $R_0\!\sim\!\sqrt{a_0}\!\ll\!L_0$. Under these conditions --- which are termed the long-wavelength, the lubrication, the slender-bar, or the shallow-water approximation in various scientific communities --- one needs to take in account only the spatial variation of fields along the main axis of the cylinder, say $x$, neglecting the variations in the transverse directions. Consequently, in this long-wavelength approximation incompressibility can be expressed as $\dot{a}\!=\!-a\,\dot{\epsilon}$ in terms of the cross-sectional area $a(x,t)$ and the extensional strain-rate $\dot{\epsilon}(x,t)\!=\!\pa_x v(x,t)$, and quasi-static linear momentum balance as $\partial_x\left(a\,\sigma\right)\!=\!0$, where $\sigma(x,t)$ is the axial stress. Note that this approximation may break down dynamically when strong localization develops. Linear elasticity, in this approximation, takes the form $\dot{\epsilon}^{el}\!=\!\dot{\sigma}/G$, where $G$ is a relevant linear elastic modulus (here it is simply the Young's modulus). Finally, we neglect the spatiotemporal variation of the temperature $T$ and focus on a single internal-state field whose evolution equation is $\dot{{\cal I}} = \dot{\epsilon}^{pl} g\left(\sigma,{\cal I}\right)$.

Taken together, our dimensionally-reduced set of equations takes the form
\begin{subequations}
\label{eq:generic_1D}
  \begin{align}
    &\pa_t{a} + v\pa_x a=-a\,\pa_x{v}\ , \label{eq:generic_1D_1}\\
    &\partial_x\!\left(a\,\sigma\right) = 0 \ , \label{eq:generic_1D_2}\\
    &\pa_t\sigma + v\pa_x \sigma=G[\pa_x{v}-\dot{\epsilon}^{pl}(\sigma,T,{\cal I})] \ , \label{eq:generic_1D_3}\\
    &\pa_t{\cal I} + v\pa_x {\cal I} = \dot{\epsilon}^{pl} g\!\left(\sigma,{\cal I}\right) \ .\label{eq:generic_1D_4}
  \end{align}
\end{subequations}
Here, Eq.~\eqref{eq:generic_1D_1} is the long-wavelength incompressible counterpart of Eq.~\eqref{eq:mass_conservation}, Eq.~\eqref{eq:generic_1D_2} is the long-wavelength quasi-static limit of Eq.~\eqref{eq:momentum_balance}, Eq.~\eqref{eq:generic_1D_3} is the scalar counterpart of Eq.~\eqref{eq:strain_rate_decomp}, and Eq.~\eqref{eq:generic_1D_4} corresponds to the scalar counterpart of Eq.~\eqref{eq:internal_variable} for a single internal-state field. In the remainder of the paper, we will study this set of equations for a long cylindrical bar of length $L(t)$ under extensional deformation generated by an imposed edge velocity $v(x\!=\!L(t)/2,t)\!=\!-v(x\!=\!-L(t)/2,t)$, where $x\!=\!0$ is the middle of the bar.

\section{Approximate time-dependent linear stability analysis}
\label{sec:LSA}

The set of Eqs.~\eqref{eq:generic_1D} admits spatially-homogeneous deformation solutions, which can be obtained once $\dot{\epsilon}^{pl}(\sigma,T,{\cal I})$ and $g\!\left(\sigma,{\cal I}\right)$ are specified. The first question we aim at addressing is the linear stability of these solutions. In many problems in physics, the spatially-homogeneous solutions are also temporally-homogeneous, i.e.~they are time-independent. This is not the case in the problem at hand, as is immediately inferred from Eq.~\eqref{eq:generic_1D_1}; the right-hand-side is finite (the initially homogeneous cross-sectional area $a$ is finite and $\pa_x{v}$ is finite because the system experiences extensional deformation). The time-dependent nature of the spatially-homogeneous solutions has serious implications for the linear stability problem.

To appreciate these implications, consider a general problem consisting of $N$ fields $f^{(i)}$, $i\!=\!1,2,\ldots N$, where the spatially-homogeneous solutions are time-dependent. We then add to these spatially-homogeneous solutions $f^{(i)}_h(t)$ --- the subscript `h' stands for spatially-homogeneous --- spatiotemporal perturbations of wavenumber $k$, leading to $f^{(i)}\!\left(x,t\right)\!=\!f^{(i)}_h\!\left(t\right)+\delta^{(i)}\!\left(t\right) e^{ikx}$. Inserting these $f^{(i)}(x,t)$'s into the complete nonlinear set of equations, in our case Eqs.~\eqref{eq:generic_1D}, and linearizing with respect to the perturbations, we obtain
\begin{equation}
\label{eq:generic_linear}
  \partial_t {\bm\delta}(t) = \bm M\!\left(t,k\right){\bm\delta}(t) \ ,
\end{equation}
where ${\bm\delta}(t)$ is a vector whose components are the time-dependent parts of the perturbation, $\delta^{(i)}\!\left(t\right)$. The crucial point is that due to the time-dependence of the spatially-homogeneous fields $f^{(i)}_h\!\left(t\right)$, the matrix $\B M$ is also time-dependent.

The time-dependence of $\B M$ has two major consequences that are missing in conventional linear stability analyses, where it is a time-independent constant. To identify these, we follow the standard procedure and diagonalize $\B M(t)$, where the dependence on $k$ is omitted here for simplicity; here the diagonalizing matrix $\B P(t)$ (which consists of the diagonalizing basis vectors) is also time-dependent, $\B P^{-1}(t) {\B M}(t) \B P(t)\!=\!\text{\bf diag}(\lambda_1(t), ..., \lambda_N(t))$, as well as the eigenvalues $\lambda_i(t)$.
Defining the vector of transformed field perturbations $\B \Delta(t)\!\equiv\!\bm{P}^{-1}(t)\B \delta(t)$, Eq.~\eqref{eq:generic_linear} transforms into
\begin{equation}
\label{eq:linear_full}
  \partial_t\B\Delta(t)\!=\!\left[\text{\bf diag}\left(\lambda_1(t),...,\lambda_N(t)\right)\!+\!\left(\partial_t\bm{P}^{-1}(t)\right)\bm{P}(t)\right]\!\B\Delta(t)\ .
\end{equation}
The two differences compared to the conventional time-independent linear stability analysis are apparent; first, the eigenvalues $\lambda_i$ do not exclusively determine the linearized evolution of the system as the time-dependence of $\B P(t)$ and its inverse play a role as well, as indicated by the second term on the right-hand-side. Second, the eigenvalues themselves are time-dependent, $\lambda_i(t)$.

To the best of our knowledge, Eq.~\eqref{eq:linear_full} does not admit a general exact solution. To proceed, we adopt a WKB-like approximation~\cite{Wentzel1926,Kramers1926,Dunham1932} in which the base vectors in $\B P(t)$ are assumed to vary slowly compared to other timescales in the problem. With this assumption, Eq.~\eqref{eq:linear_full} can be approximated as $\partial_t\B\Delta(t)\!\simeq\!\text{\bf diag}(\lambda_1(t),...,\lambda_N(t))\B\Delta(t)$. Solving this equation and transforming back to the original fields, we obtain
\begin{equation}
\label{eq:linear_solution_generic}
  \B\delta\left(\Delta t, t_0\right)\!\simeq\! \sum_{i}^{N}\Delta_i\!\left(t_0\right) {\B{\mathcal{E}}}_i\!\left(t_0\right) \exp\!\left[{\int_{t_0}^{t_0+\Delta t}\!\!\lambda_i\!\left(t'\right)dt'}\right]\ ,
\end{equation}
where ${\B{\mathcal{E}}}_{i}\!\left(t_0\right)$ are the eigenvectors of $\bm M\!\left(t\right)$ at time $t\!=\!t_0$, $\Delta_{i}\!\left(t_0\right)$ are constants obtained from $\B\delta\!\left(t_0\right)$ (the perturbation at time $t_0$) according to $\B \Delta(t_0)\!\equiv\!\bm{P}^{-1}(t_0)\B \delta(t_0)$ and $\Delta{t}$ is the time measured from $t_0$. While Eq.~\eqref{eq:linear_solution_generic} is an exact solution to the approximated equation, the quality of the approximation to the exact Eq.~\eqref{eq:linear_full} and its dependence on $t_0$ and the time interval $\Delta{t}$ are not a priori known. Finally, we simplify Eq.~\eqref{eq:linear_solution_generic} by further assuming that the largest eigenvalue, $\lambda_+(t)$, dominates the sum and that we can set $t_0\!=\!0$ and $\Delta{t}\!=\!t$, resulting in
\begin{equation}
\label{eq:linear_solution_approx}
\B\delta\!\left(t\right) \simeq \Delta_+\!\left(0\right) {\B{\mathcal{E}}}_+\!\left(0\right) \exp\!\left[{\int_{0}^{t}\!\!\lambda_+\!\left(t'\right)dt'}\right]\ .
\end{equation}
Here ${\B{\mathcal{E}}}_{+}\!\left(0\right)$ is the eigenvector corresponding to the largest eigenvalue $\lambda_+$ at $t\!=\!0$ and $\Delta_+\!\left(0\right)$ is the corresponding constant.

We now aim at applying the approximated expression for the time-dependent linear stability analysis in Eq.~\eqref{eq:linear_solution_approx} to the necking problem, as formulated in Eqs.~\eqref{eq:generic_1D}. First, we note that the convective derivative terms in Eqs.~\eqref{eq:generic_1D}, $v\pa_x \bullet$, vanish identically for the spatially-homogeneous dynamics or are small within the linear perturbation regime around the spatially-homogeneous state, hence they are omitted from the analysis (though they are included in any direct numerical solutions of Eqs.~\eqref{eq:generic_1D} below). In their absence, Eqs.~\eqref{eq:generic_1D} become independent of the wavenumber of perturbations $k$. Note in this context that the velocity gradient $\pa_x{v}$ is finite in the absence of perturbations and its value in this case is determined by the applied velocity. In particular, in the spatially-homogeneous state, $\pa_x{v}$ is identical to its spatial average $\avg{\pa_x v}\!\equiv\![L(t)]^{-1}\!\int_{-\tfrac{1}{2}L(t)}^{\tfrac{1}{2}L(t)}\pa_x{v}\,dx\!=\!\dot{L}(t)/L(t)\!=\!\dot{\epsilon}_{h}(t)$, which is the imposed Hencky strain-rate. In the presence of spatial inhomogeneity, $\dot{\epsilon}_{h}(t)$ is in fact the {\em average} imposed Hencky strain-rate, $\dot{\bar{\epsilon}}(t)\!=\!\dot{\epsilon}_{h}(t)$.

Throughout this paper the applied boundary conditions correspond to a constant, time-independent Hencky strain-rate $\dot{\bar{\epsilon}}$, though the salient features of the results do not change for other boundary conditions, e.g.~constant velocity. For the spatially-homogeneous dynamics $\dot{\epsilon}_{h}\!=\!\dot{\bar{\epsilon}}$ is just a parameter and the time-dependent solutions for $a_h(t)$, $\sigma_h(t)$ and ${\cal I}_h(t)$ can be obtained. We then introduce perturbations to all $4$ fields, $\delta\dot{\epsilon}(t)$, $\delta{a}(t)$, $\delta\sigma(t)$ and $\delta{\cal I}(t)$ and linearize Eqs.~\eqref{eq:generic_1D} around the spatially-homogeneous state (recall that the convective derivative terms are omitted and $\pa_x{v}\!=\!\dot\epsilon$ in Eqs.~\eqref{eq:generic_1D}). To proceed, we eliminate $\delta\dot{\epsilon}$ between Eqs.~\eqref{eq:generic_1D_1} and~\eqref{eq:generic_1D_3}, and express $\delta\sigma$ in terms of $\delta{a}$ using Eq.~\eqref{eq:generic_1D_2} (which implies $\sigma_h\delta a \!=\!-a_h\delta \sigma$ to linear order).

We are then left with two fields, i.e.~$\B\delta\!=\!\left(\delta a,\delta {\cal I}\right)^{T}$, and the matrix $\B M$ in Eq.~\eqref{eq:generic_linear} takes the approximate form
\begin{equation}
\label{eq:M_generic_hard_geometry}
  \bm{M}\!\left(t\right)\simeq\left(\begin{array}{cc}
\bigstrut\sigma\partial_{\sigma}\dot{\epsilon}^{pl} - \dot{\epsilon}^{pl} & \bigstrut-a\partial_{{\cal I}}\dot{\epsilon}^{pl}\\
\bigstrut-\frac{\sigma}{a}\partial_{\sigma}\dot{\cal I} & \bigstrut\partial_{{\cal I}}\dot{\cal I}
\end{array}\right)\ ,
\end{equation}
where the subscript `h' is omitted hereafter. In Eq.~\eqref{eq:M_generic_hard_geometry} we used the fact that the stress is typically much smaller than the elastic modulus, $\sigma\!\ll\!G$. The time-dependent eigenvalues can be readily obtained as $\lambda(t)\!=\!\tfrac{1}{2}\text{tr}\bm{M}\!\left(1\!\pm\!\sqrt{1 - 4\,\text{det}\bm{M}/(\text{tr}\bm{M})^2}\right)$. For constitutive relations for which $\text{det}\bm{M}\!\ll\!(\text{tr}\bm{M})^2$, the largest eigenvalue --- the most important physical quantity of interest in the linearized analysis --- attains a simple analytical expression
\begin{equation}
\label{eq:lambda_+_generic}
  \lambda_+\!\left(t\right)\simeq \text{tr}\bm{M} = \sigma\partial_{\sigma}\dot{\epsilon}^{pl} - \dot{\epsilon}^{pl} + \partial_{{\cal I}}\dot{{\cal I}}\ ,
\end{equation}
in case it is positive. In case the latter expression is negative, the largest eigenvalue is $\lambda_+\!(t)\!\simeq\!0$. Consequently, while we hereafter refer to Eq.~\eqref{eq:lambda_+_generic} as the largest eigenvalue, it should be understood that it is valid only if it is positive; otherwise, $\lambda_+\!(t)\!=\!0$ is used.

\section{The largest eigenvalue and the onset of necking}
\label{sec:largest_eigenvalue}

The previous section culminated with an analytical approximation, cf.~Eq.~\eqref{eq:lambda_+_generic}, for the largest eigenvalue $\lambda_+(t)$ in the time-dependent linear stability analysis for the necking problem. Let us briefly discuss the physical meaning of the different contributions to $\lambda_+(t)$. The appearance of the stress $\sigma$ in the first term on the right-hand-side is a direct consequence of the coupling between the system's geometry and its mechanical response, resulting in stress amplification due to shape perturbations. The stress $\sigma$ multiplies the variation of the plastic strain-rate with the stress,  $\partial_{\sigma}\dot{\epsilon}^{pl}$, at constant internal structure $\mathcal{I}$ and temperature $T$. As the stress is the driving force for plastic deformation, we expect $\dot{\epsilon}^{pl}$ to generically increase with increasing $\sigma$, at constant internal structure $\mathcal{I}$ and temperature $T$. Consequently, the first term on the right-hand-side of Eq.~\eqref{eq:lambda_+_generic} is positive, $\sigma\partial_{\sigma}\dot{\epsilon}^{pl}\!>\!0$, i.e.~it generically promotes a necking instability.

The second term on the right-hand-side of Eq.~\eqref{eq:lambda_+_generic} is the (extensional, hence positive) plastic strain-rate $\dot{\epsilon}^{pl}$ with a minus sign, i.e.~this contribution is intrinsically stabilizing. That is, homogeneous plastic deformation, by itself, tends to limit the growth of shape perturbations. We note that the ratio of the second to first terms on the right-hand-side of Eq.~\eqref{eq:lambda_+_generic} is the dimensionless strain-rate sensitivity $m^{-1}\!\equiv\!\pa\log(\sigma)/\pa\log(\dot{\epsilon}^{pl})$, which is small for many materials~\cite{Estrin1984,Hutchinson1977}. Hence we expect the first term to dominate the second one in many cases.

The third contribution to $\lambda_+(t)$ in Eq.~\eqref{eq:lambda_+_generic}, $\partial_{{\cal I}}\dot{{\cal I}}$, concerns the evolution of the material's structure --- described by the coarse-grained internal-state field ${\cal I}$ --- which inevitably accompanies plastic deformation. To discuss the physical meaning of this contribution one needs to be a bit more explicit about the physical nature of the scalar internal-state field ${\cal I}$. To that aim, it would be natural to assume that ${\cal I}$ corresponds to the density of plasticity carriers in the material, for example dislocations in crystalline materials~\cite{Mecking1981,Estrin1984,Follansbee1988,Kocks2003} or Shear-Transformation-Zones (STZs) in amorphous materials~\cite{Spaepen1977,Argon1979,Falk1997}. With this physical picture in mind, $\partial_{{\cal I}}\dot{{\cal I}}$ represents the variation of the rate-of-change of the density of plasticity carriers with their density. For materials in which locked-in plasticity carriers hamper subsequent plastic deformation, we expect $\partial_{{\cal I}}\dot{{\cal I}}\!<\!0$. This is the case for strain-hardening materials such as metals, for which this contribution is stabilizing. On the other hand, for strain-softening materials such as bulk metallic glasses~\cite{Gilman1975,Schuh2007,Chen2008,Wang2009,Trexler2010,Cheng2011a,Wang2012,Egami2013,Hufnagel2016a}, the activation of plasticity carriers facilitates subsequent plastic deformation and we have $\partial_{{\cal I}}\dot{{\cal I}}\!>\!0$. Consequently, and not quite surprisingly, Eq.~\eqref{eq:lambda_+_generic} predicts that strain-softening materials are more prone to necking instabilities, while in strain-hardening materials necking instabilities may be at least delayed to larger strains.

The largest eigenvalue in Eq.~\eqref{eq:lambda_+_generic} can be used to derive the {\em onset} conditions for necking, i.e.~$\lambda_+(t)\!>\!0$. In the next section we will argue that more significant, physically meaningful and predictive information about necking can be extracted from the time-dependent eigenvalue $\lambda_+(t)$. In the meantime, however, we do consider the onset condition $\lambda_+(t)\!>\!0$ and compare it to conventional onset conditions available in the literature. Apparently the first, and quite extensively used, criterion for the onset of necking is Consid\`ere's criterion $d\sigma/d{\epsilon^{pl}}\!<\!\sigma$, which assumes that the stress $\sigma$ can be expressed solely as a function of the plastic strain $\epsilon^{pl}$. As is evident from the discussion in Sect.~\ref{sec:formulation}, this is not a physically realistic formulation --- the plastic strain itself is not a legitimate independent physical field, but rather can be calculated from a time-integral over the plastic strain-rate $\dot\epsilon^{pl}$. As such, we cannot directly compare our onset criterion $\lambda_+(t)\= \sigma\partial_{\sigma}\dot{\epsilon}^{pl}-\dot{\epsilon}^{pl}+\partial_{{\cal I}}\dot{{\cal I}}\!>\!0$ to Consid\`ere's criterion.

An extensively used generalization of Consid\`ere's criterion, taking into account the rate-dependence of plastic deformation --- i.e.~considering a relation of the form $\sigma({\epsilon}^{pl}, \dot{\epsilon}^{pl})$ --- has been proposed by Hart~\cite{Hart1967}. The Hart criterion reads
\begin{equation}
\frac{1}{m} + \frac{\pa\sigma}{\sigma \pa{\epsilon^{pl}}} < 1 \ ,
\label{eq:Hart}
\end{equation}
where the strain-rate sensitivity $m^{-1}$ has been defined above. By dividing by $\sigma\pa_\sigma \dot\epsilon^{pl}\!\!>\!0$ and rearranging, our onset criterion $\lambda_+(t)\= \sigma\partial_{\sigma}\dot{\epsilon}^{pl}\!-\!\dot{\epsilon}^{pl}\!+\!\partial_{{\cal I}}\dot{{\cal I}}\!>\!0$ can be cast in a form somewhat reminiscent of Eq.~\eqref{eq:Hart}
\begin{equation}
 \frac{1}{m}-\frac{\pa_{\cal I} \dot{\cal I}}{\sigma\pa_\sigma \dot\epsilon^{pl}} < 1 \ .
\label{eq:Hart-like}
\end{equation}
Comparing Eq.~\eqref{eq:Hart-like} to Eq.~\eqref{eq:Hart} we observe that $-\sigma^{-1}\pa_{\cal I} \dot{\cal I}/\pa_\sigma \dot\epsilon^{pl}$ appears to play the role of the dimensionless strain-hardening/softening coefficient $\sigma^{-1}\pa\sigma/\pa{\epsilon^{pl}}$ (which is positive for strain-hardening materials and negative for strain-softening materials). Indeed, as discussed above, for strain-hardening materials we expect $\pa_{\cal I} \dot{\cal I}\!\!<\!0$, i.e.~$-\sigma^{-1}\pa_{\cal I} \dot{\cal I}/\pa_\sigma \dot\epsilon^{pl}\!\!>\!0$ and for strain-softening materials we expect $\pa_{\cal I}\dot{\cal I}\!\!>\!0$, i.e.~$-\sigma^{-1}\pa_{\cal I} \dot{\cal I}/\pa_\sigma \dot\epsilon^{pl}\!\!<\!0$. While this analogy might appear suggestive, there is no fundamental reason to expect $-\sigma^{-1}\pa_{\cal I} \dot{\cal I}/\pa_\sigma \dot\epsilon^{pl}$ to correspond to the measured strain-hardening/softening coefficients extracted from stress-strain curves. Instead, we believe that Eq.~\eqref{eq:Hart-like} offers a more physically sound criterion for the onset of necking, going beyond the criteria that are available in the literature.

\section{Strong localization criterion}
\label{sec:localization}

The onset criterion in Eq.~\eqref{eq:Hart-like} predicts the time, or equivalently the strain, at which shape perturbations start to grow, $\lambda_+(t)\!>\!0$. As such, it does not provide information about the evolution of the neck and in particular about the emergence of macroscopic shape localization. The approximate time-dependent linear stability analysis developed in Sect.~\ref{sec:LSA} is, in fact, capable of providing more extensive and quantitative predictions about necking dynamics, as we show next. Combining the results in Eqs.~\eqref{eq:linear_solution_approx} and~\eqref{eq:lambda_+_generic}, we obtain the time-evolution of the cross-sectional area (shape) perturbation as
\begin{equation}
\label{eq:linear_solution_approx_a}
\delta{a}(t)\!\simeq\!\Delta_+(0) {{\mathcal{E}}}_+\!^{(a)}(0) \exp\!\left[\int_{0}^{t}\!\!\left(\sigma\partial_{\sigma}\dot{\epsilon}^{pl}\!-\!\dot{\epsilon}^{pl}\!+\!\partial_{{\cal I}}\dot{{\cal I}}\right)dt'\right]\ ,
\end{equation}
where ${{\mathcal{E}}}_+\!^{(a)}$ is the component of the vector ${\B{\mathcal{E}}}_+$ that corresponds to $\delta{a}$ and all of the functions in the integrand are understood to depend on the time $t'$.

We now aim at using $\delta{a}(t)$ of Eq.~\eqref{eq:linear_solution_approx_a}, obtained from the approximated linear analysis, to predict the onset of strong, nonlinear shape localization. Such a strong localization will be macroscopically manifested and might eventually lead to catastrophic failure (e.g. by shear banding, cavitation, crack propagation or the like), which goes beyond the scope of this paper. Strong localization is expected to initiate once the perturbation $\delta{a}(t)$ becomes non-negligible compared to the homogeneous cross-sectional area $a_h(t)\=a_0\exp(-\dot{\bar{\epsilon}}\,t)$, which is the spatially-homogeneous solution of Eq.~\eqref{eq:generic_1D_1} for a constant $\dot{\bar{\epsilon}}$. That is, we are looking for the time in the necking dynamics at which $\delta{a}(t)/a_h(t)\=\kappa\!\sim\!{\cal O}(10^{-1})$. While the exact value of $\kappa$, i.e.~the ratio of the perturbation to the homogeneous solution at which non-negligible nonlinearities set in, is not a priori known, we show below that the results are only weakly dependent on $\kappa$ in this range. Note also that both the growing perturbation after the onset of necking and the exponential time decay of the homogeneous solution tend to increase the relative magnitude of the perturbation in the ratio $\delta{a}(t)/a_h(t)$.

Applying this criterion, the onset of strong necking-mediated localization is predicted to occur at $t\=t_\ell$, which is a solution of the following integral equation
\begin{equation}
\label{eq:strong_localization}
\int_{0}^{t_\ell}\!\!\left(\sigma\partial_{\sigma}\dot{\epsilon}^{pl}+\partial_{{\cal I}}\dot{{\cal I}}\right)dt' \simeq \log\left(\frac{\kappa}{\zeta}\right) \ ,
\end{equation}
where the negligibly small average elastic strain-rate $\dot{\bar{\epsilon}}-\dot{\epsilon}^{pl}$ has been omitted from the integrand and we defined $\zeta\!\equiv\!\Delta_+(0){\mathcal{E}}_+\!^{(a)}(0)/a_0$. To make use of Eq.~\eqref{eq:strong_localization} we need to evaluate the right-hand-side, which involves $\kappa\!\sim\!{\cal O}(10^{-1})$ and the relative magnitude of the initial ($t\=0$) perturbation. The exact value of $\kappa\!\sim\!{\cal O}(10^{-1})$ does not make a significant quantitative difference due to the logarithmic dependence. $\zeta$ is determined either by the prescribed perturbation in a deterministic calculation or by the typical noise level in a realistic physical situation; as such, $\zeta$ in Eq.~\eqref{eq:strong_localization} is interpreted as a characteristic (relative) amplitude of perturbations, being more general than its formal definition given above in the framework of the linear stability analysis.

We believe that the prediction in Eq.~\eqref{eq:strong_localization} for the emergence of strong localization is more physically significant and relevant compared to the onset of necking predicted in Eq.~\eqref{eq:Hart-like}. While both are based on a linearized analysis, the former in fact provides an estimate for the onset of nonlinearities. Indeed, as has been mentioned in Sect.~\ref{sec:intro}, several works highlighted the limitations of linearized onset criteria, trying to circumvent them by studying constitutive relations that allow some nonlinear progress and invoking several simplifying geometrical assumptions. In particular, Hutchinson and Neale~\cite{Hutchinson1977} performed such a nonlinear analysis for rate-dependent viscoplastic materials and Fressengeas and Molinari~\cite{Fressengeas1985} extended the analysis to include also thermal effects on the localization process, though some other simplifying assumptions have been made (e.g.~a constant applied force has been assumed). While these works have provided some analytical insight into the nonlinear evolution of necking, they seem to lack the degree of generality of Eq.~\eqref{eq:strong_localization}, both in terms of the adopted constitutive framework and in terms of the additional simplifying assumptions invoked. Note also that~\cite{Hutchinson1977, Fressengeas1985}, like the present work, invoke the long-wavelength (lubrication/slender bar) approximation, which is anyway likely to break down when strong nonlinearities set in.

In the reminder of this paper we aim at quantitatively testing Eq.~\eqref{eq:strong_localization} for two widely different elasto-viscoplastic constitutive relations, describing different classes of materials, against fully nonlinear numerical solutions of Eqs.~\eqref{eq:generic_1D}. Before discussing these quantitative analyses in detail in the next sections, we would like to briefly demonstrate how such analyses are performed. To assess the quality of the proposed criterion, we need to obtain fully nonlinear solutions of Eqs.~\eqref{eq:generic_1D}. Such numerical solutions, once $\dot{\epsilon}^{pl}(\sigma,T,{\cal I})$ and $g(\sigma, {\cal I})$ are specified, are obtained by first transforming the equations to the Lagrangian frame of reference, in which the integration domain is time-independent. Then the equations are solved by a conventional technique, see \hyperref[appendix]{Appendix} for details.
\begin{figure}[h!]
\includegraphics[width=0.48\textwidth]{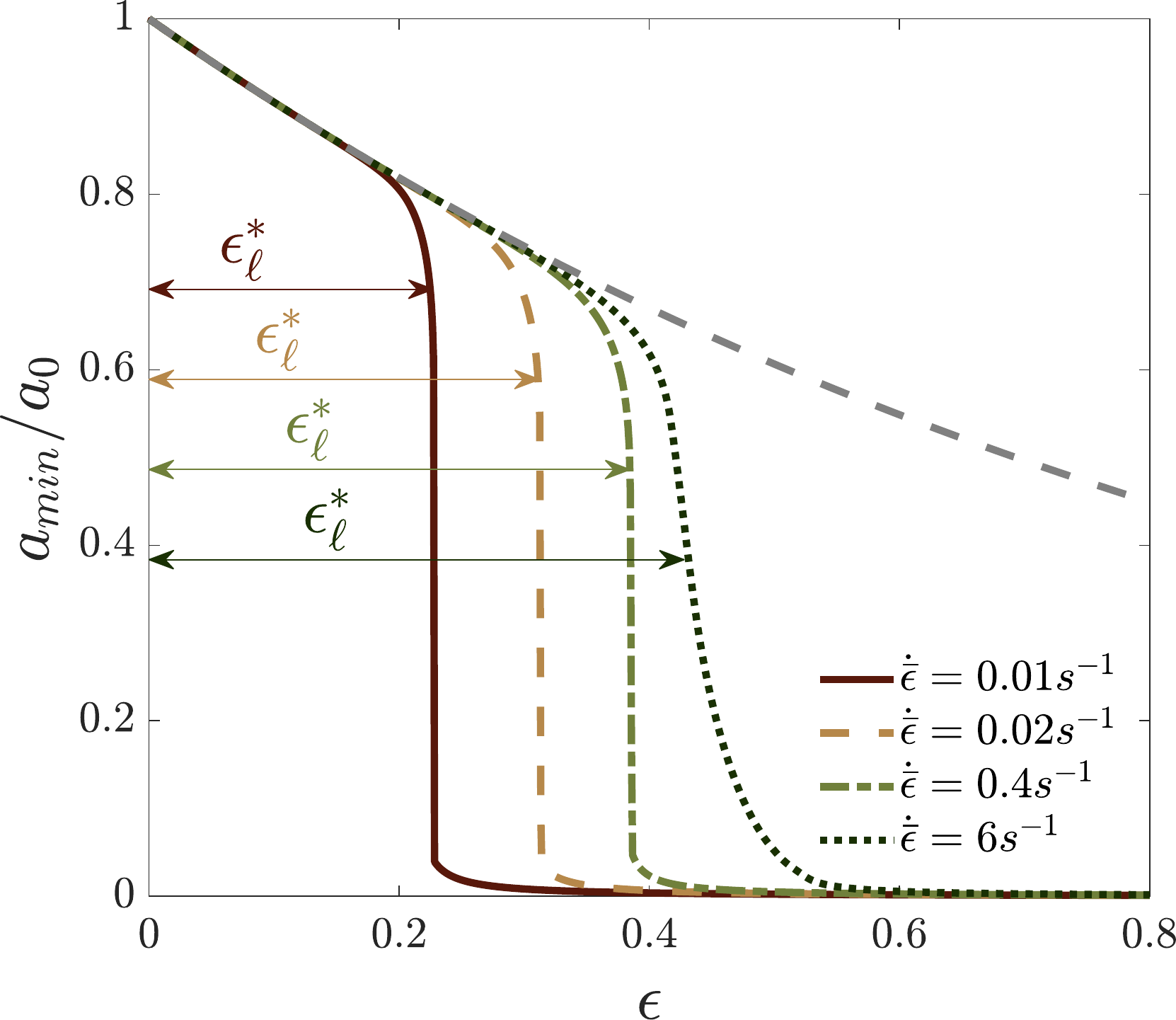}
\caption{The minimal normalized cross-sectional area $a_{min}/a_0$, as a function of the accumulated Hencky strain $\epsilon$ (measured from the onset of plasticity), for various applied strain-rates $\dot{\bar{\epsilon}}$ (see legend). Minimal cross-sections were obtained from fully nonlinear numerical solutions of Eqs.~\eqref{eq:generic_1D} for the amorphous plasticity model discussed in Sect.~\ref{sec:STZ}, with an initial inhomogeneity imposed on the internal-state field. The solution in which spatial homogeneity is imposed is added for reference (dashed gray line), as it is strain-rate independent. The curves corresponding to the perturbed solutions significantly deviate from the spatially-homogeneous solution at characteristic strains marked by the arrows and denoted by $\epsilon^*_\ell$.}
\label{fig:figure2}
\end{figure}

The system is driven by an applied Hencky strain-rate $\dot{\bar{\epsilon}}$ and the spatially-homogeneous initial conditions are supplemented with a perturbation $\delta{\cal I}(t\=0)\=\zeta\,{\cal I}_0\cos(2\pi x/L_0)$ of the internal-state field ${\cal I}$, where ${\cal I}_0$ is its homogeneous initial value. In addition to tracking the time-evolution of the perturbed system, we also solve Eqs.~\eqref{eq:generic_1D} by enforcing spatial homogeneity; this time-dependent spatially-homogeneous solution serves as a reference for evaluating the relative importance of spatial inhomogeneity. For convenience, we use the Hencky strain $\epsilon\!\equiv\!\dot{\bar{\epsilon}}\,t$ as the independent variable instead of the time $t$, which also renders the homogeneous cross-sectional area evolution, $a_h(\epsilon)\=a_0\exp(-\epsilon)$, strain-rate independent.

In Fig.~\ref{fig:figure2} we present an example of such solutions using a constitutive relation to be discussed in detail in Sect.~\ref{sec:STZ} and a relative perturbation amplitude $\zeta\=10^{-6}$. To quantify the evolution of the system, we plot the spatial minimum of the cross-sectional area, $a_{min}(\epsilon)\!\equiv\!\min\left[a(x,\epsilon)\right]$, for various applied strain-rates $\dot{\bar{\epsilon}}$, together with the homogeneous solution $a_h(\epsilon)$. We observe a generic behavior in which the perturbed solutions appear indistinguishable from the homogeneous solution (this point will be further discussed below) over a certain range of extensional strains $\epsilon$, followed by a rather abrupt drop in the minimal cross-sectional area of the perturbed solutions relative to the homogeneous one, marking the onset of strong localization. The latter, denoted by $\epsilon^*_\ell$ and schematically marked on the figure, exhibits rate-dependence.

\begin{figure}[h!]
\includegraphics[width=0.48\textwidth]{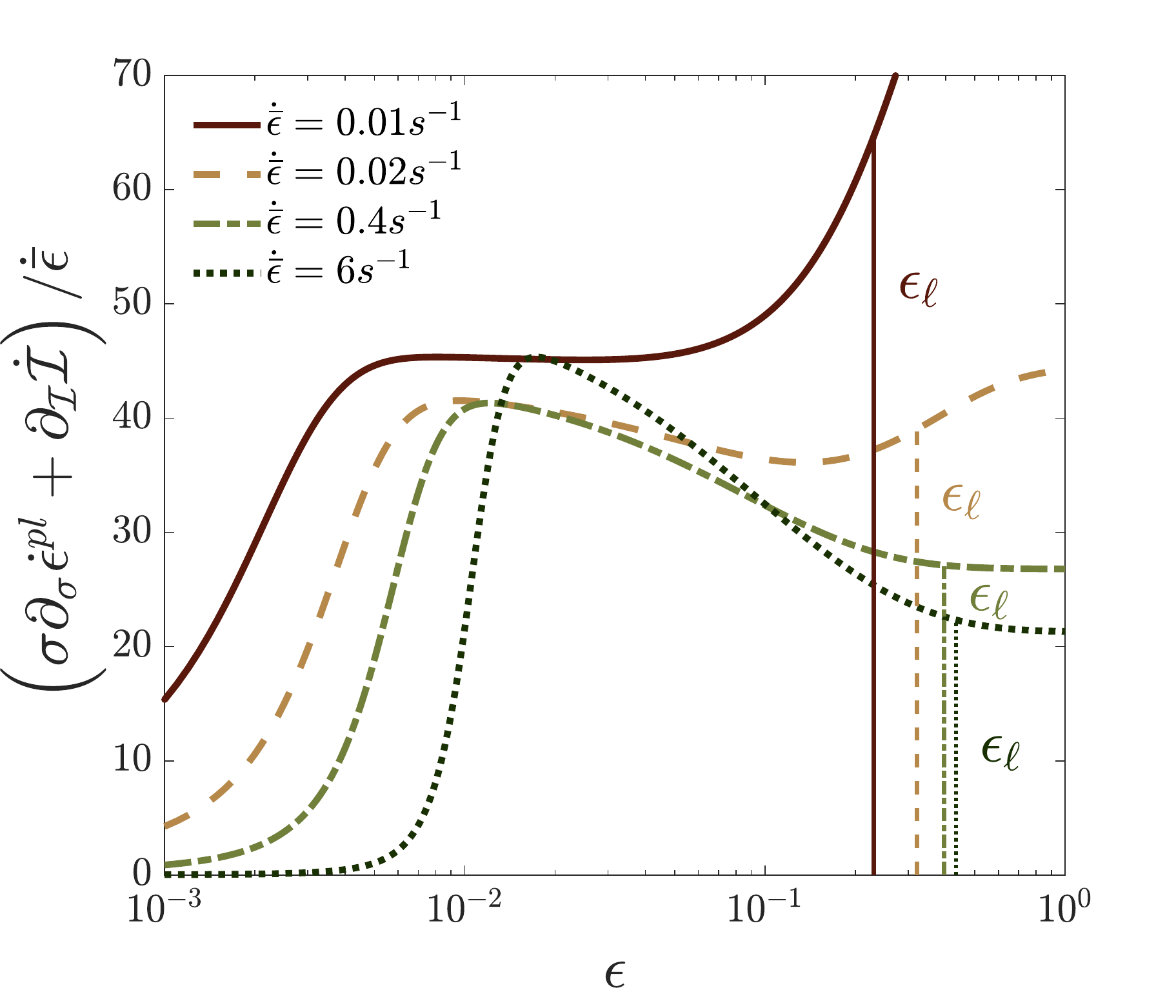}
\caption{$(\sigma\partial_{\sigma}\dot{\epsilon}^{pl}+\partial_{{\cal I}}\dot{{\cal I}})/\dot{\bar{\epsilon}}$ (the integrand of Eq.~\eqref{eq:strong_localization} normalized by the applied strain-rate $\dot{\bar{\epsilon}}$), as a function of the accumulated Hencky strain $\epsilon$ (measured from the onset of plasticity) for various strain-rates (for homogeneous solutions with the parameters of Fig.~\ref{fig:figure2}). The solutions of Eq.~\eqref{eq:strong_localization}, $\epsilon_\ell\!=\!\epsilon(t_\ell)$, are marked by the vertical lines.}
\label{fig:figure3}
\end{figure}

As explained above, the ultimate goal of the developed theory is to quantitatively predict the strong localization strain $\epsilon^*_\ell$ using Eq.~\eqref{eq:strong_localization}, which results in the prediction $\epsilon_\ell\!\equiv\!\epsilon(t_\ell)$. In Fig.~\ref{fig:figure3} we plot the integrand of the integral relation in Eq.~\eqref{eq:strong_localization} (normalized by the applied strain-rate $\dot{\bar{\epsilon}}$) as a function of the strain $\epsilon$ (instead of the time $t$) for homogeneous solutions (parameters as in Fig.~\ref{fig:figure2}). The first important observation we make is that the integrand --- which is closely related to the largest eigenvalue of Eq.~\eqref{eq:lambda_+_generic} --- is positive and increases with $\epsilon$ in the very same strain interval in which the perturbed systems appear to be indistinguishable from the spatially-homogeneous system in Fig.~\ref{fig:figure2}. This observation clearly demonstrates that while the onset condition is important, it provides only partial information about the macroscopic development of instability and the onset of strong localization. Obviously, the relation between the onset of instability and the onset of strong localization depends on the typical magnitude of perturbations quantified by $\zeta$ (e.g.~set by the noise level in the system), as is clear from Eq.~\eqref{eq:strong_localization}.

The onset of strong localization can be obtained by integrating the integrand shown in Fig.~\ref{fig:figure3} until a value determined by $\kappa$ on the right-hand-side of Eq.~\eqref{eq:strong_localization} is reached. In Fig.~\ref{fig:figure3} we used vertical lines to mark the strain values $\epsilon_\ell$ for which Eq.~\eqref{eq:strong_localization} is satisfied for $\kappa\=0.15$. In what follows, we compare the predictions for the strong localization strain $\epsilon_\ell$ to the strong localization strain $\epsilon^*_\ell$ obtained directly from fully nonlinear solutions for two widely different elasto-viscoplastic constitutive relations, describing different classes of materials, and for various physical conditions and material parameters.

\section{Application I: Amorphous materials and the STZ model}
\label{sec:STZ}

The analysis up to now remained fairly general, and as such could be applied to any constitutive law that involves a single internal-state field (the generalization to more internal-state fields is straightforward, but will be more mathematically involved). We derived an instability onset criterion as the zero-crossing in time of the largest time-dependent eigenvalue in Eq.~\eqref{eq:lambda_+_generic}, discussed its relations to existing criteria, and obtained a strong localization criterion in Eq.~\eqref{eq:strong_localization}. In order to quantitatively test these criteria, one should specify a constitutive law, i.e.~$\dot{\epsilon}^{pl}(\sigma,T,{\cal I})$ and $g(\sigma, {\cal I})$. In this and the following sections, we consider two widely different constitutive laws --- one for amorphous/glassy materials and one for crystalline/polycrystalline materials --- and use them to quantitatively test our predictions.

We start by considering amorphous/glassy materials which lack the long-range order of crystalline materials. The elasto-viscoplastic response of amorphous/glassy materials has attracted a lot of interest in the last few decades~\cite{Falk2010,Trexler2010,Li2016, Rodney2011, Hufnagel2016, RevModPhys2017}. It is now widely accepted that plastic deformation in these materials is mediated by spatially-localized immobile rearrangements of predominantly shear nature ---  Shear-Transformation-Zones (STZs) ---, which are qualitatively different from dislocations in crystalline materials. While various fundamental questions about the nature and properties of plastic deformation in amorphous/glassy materials are still debated and intensively investigated, a few quite successful coarse-grained phenomenological models have emerged in the literature such as the Soft Glassy Rheology (SGR) model~\cite{Sollich1997,Cates2004,Fielding2009} and the Shear-Transformation-Zone (STZ) model~\cite{Falk1997,Bouchbinder2009a,Bouchbinder2009b,Bouchbinder2009c,Falk2010}.

We will consider here the STZ model. The reason for this choice is that this model includes an internal-state field and is most suitable for studying transient, far from steady-state elasto-viscoplastic deformation of amorphous/glassy materials. Moreover, this relatively simple model, which is formulated within a nonequilibrium thermodynamic framework, has been shown to capture various salient features of the elasto-viscoplastic deformation of amorphous/glassy materials~\cite{Falk1997, Langer2001,Bouchbinder2007a,Bouchbinder2007b,Langer2004,Langer2006,Langer2007,Langer2008,Bouchbinder2009a,Bouchbinder2009b,Bouchbinder2009c,Falk2010,Langer2010,Bouchbinder2011,Bouchbinder2011b,Langer2012,Langer2015}.
In particular, it has been shown to exhibit shear-banding~\cite{Manning2007,Manning2009}, necking~\cite{Eastgate2003,Eastgate2005}, and to predict a brittle-to-ductile-like transition in the fracture toughness of glasses as a function of the their preparation protocol~\cite{Rycroft2012,Vasoya2016}.

Plastic deformation in amorphous/glassy materials results from the macroscopic accumulation of microscopic irreversible strain at the cores of STZs, which corresponds to a transition between the STZ internal states that are separated by a barrier. The coarse-grained plastic strain-rate in the framework of the STZ model, adapted to our long-wavelength approximation, takes the form~\cite{Falk2010}
\begin{subequations}
\label{eq:STZ_plastic}
\begin{eqnarray}
  \dot{\epsilon}^{pl}(\sigma,T,\Lambda) &=& \Lambda(\chi) \,\,\tau_0^{-1}\mathcal{C}(\sigma,T) \,\left(1-\frac{\sigma_y}{\sigma}\right)\ ,\label{eq:STZ_plasticA}\\
   \Lambda(\chi) &=& \exp\!\left({-\frac{e_z}{k_B\chi}}\right) \ .\label{eq:STZ_plasticB}
\end{eqnarray}
\end{subequations}
In Eq.~\eqref{eq:STZ_plasticA}, $\Lambda(\chi)$ is the density of STZs, $\tau_0^{-1}\mathcal{C}(\sigma,T)$ is the stress $\sigma$- and temperature $T$-dependent barrier transition rate between the STZ internal states ($\tau_0$ is a microscopic vibration time-scale) and $\sigma_y$ is the yield stress. We consider relatively low temperatures $T$ such that no spontaneously thermally-activated creation/annihilation of STZs takes place and the yield stress $\sigma_y$ represents a rather sharp transition. Consequently, Eq.~\eqref{eq:STZ_plasticA} describes the post-yielding regime $\sigma\!>\!\sigma_y$, where as $\dot{\epsilon}^{pl}\=0$ for $\sigma\!\le\!\sigma_y$. In addition, heat generation is assumed to be sufficiently slow such that the temperature $T$ remains nearly equilibrated with the heat reservoir at any time.

The dimensionless transition rate is given by a conventional stress-biased thermal activation expression $\mathcal{C}(\sigma,T)\=\exp\!\left({-\frac{\Delta}{k_B T}}\right)\cosh\!\left(\frac{\bar{\Omega}\,\sigma}{k_B T}\right)$, where $\Delta$ is a typical activation barrier and $\bar{\Omega}$ is the product of a typical microscopic STZ volume and a typical local strain contributed by any STZ transition. If the stress becomes very large, e.g.~near the tip of a crack or inside a narrow neck, activation barriers are expected to be completely washed out and other dissipative mechanisms kick in. In particular, when the stress satisfies $\sigma\!>\!\Delta/\bar{\Omega}$, we take the stress-biased thermal activation expression to cross over to the linear relation $\mathcal{C}(\sigma,T)\=\bar{\Omega}\,\sigma/(2\Delta)$.

The density of STZs, $\Lambda$, is given in Eq.~\eqref{eq:STZ_plasticB} in terms of a Boltzmann-like factor in which $\chi$ is an effective temperature and $e_z$ is the typical STZ formation energy ($k_B$ is Boltzmann's constant)~\cite{Falk2010}. The effective temperature $\chi$ is a thermodynamic temperature that describes the slow configurational degrees of freedom of an amorphous/glassy material which fell out of equilibrium with the vibrational degrees of freedom described by $T$~\cite{Falk2010}. Within the two-temperature thermodynamic theory of deforming glasses, $\chi$ evolves according to the following effective heat equation
\begin{equation}
\label{eq:STZ_IV}
  \dot{\chi}\!=\!\frac{2 \sigma \dot{\epsilon}^{pl}}{c_0 \sigma_y}\left(\chi_{\infty}-\chi\right) \ ,
\end{equation}
where $c_0$ is the effective heat capacity, $\chi_{\infty}$ is the steady-state effective temperature (the initial value of $\chi$, $\chi_0$, typically satisfies $\chi_0\!<\!\chi_\infty$) and no $\chi$-diffusion is taken into account.

Equation~\eqref{eq:STZ_IV} suggests that $\chi$ is driven by the plastic dissipation power $\sigma \dot{\epsilon}^{pl}$ and that the fraction of the dissipation power that is stored in the material's structure (``stored energy of cold work'') is inversely proportional to the yield stress $\sigma_y$ (the complementary part is irreversibly transferred into the heat reservoir). The increase of $\chi$ with plastic deformation, which leads to enhanced plastic deformation, represents a ``flow induces flow'' process, characteristic of strain-softening materials. The initial value of $\chi$, $\chi_0$, depends on the history of the glass, e.g.~the cooling-rate through the glass transition or the aging time in the vicinity of the glass temperature, and hence describes the initial non-equilibrium state of the glass. As $\Lambda$ is uniquely related to $\chi$ through Eq.~\eqref{eq:STZ_plasticB}, Eq.~\eqref{eq:STZ_IV} serves as the relevant $\dot{\cal I}$ equation and the constitutive law is fully specified.
\begin{figure*}
\centering
\includegraphics[width=0.88\textwidth]{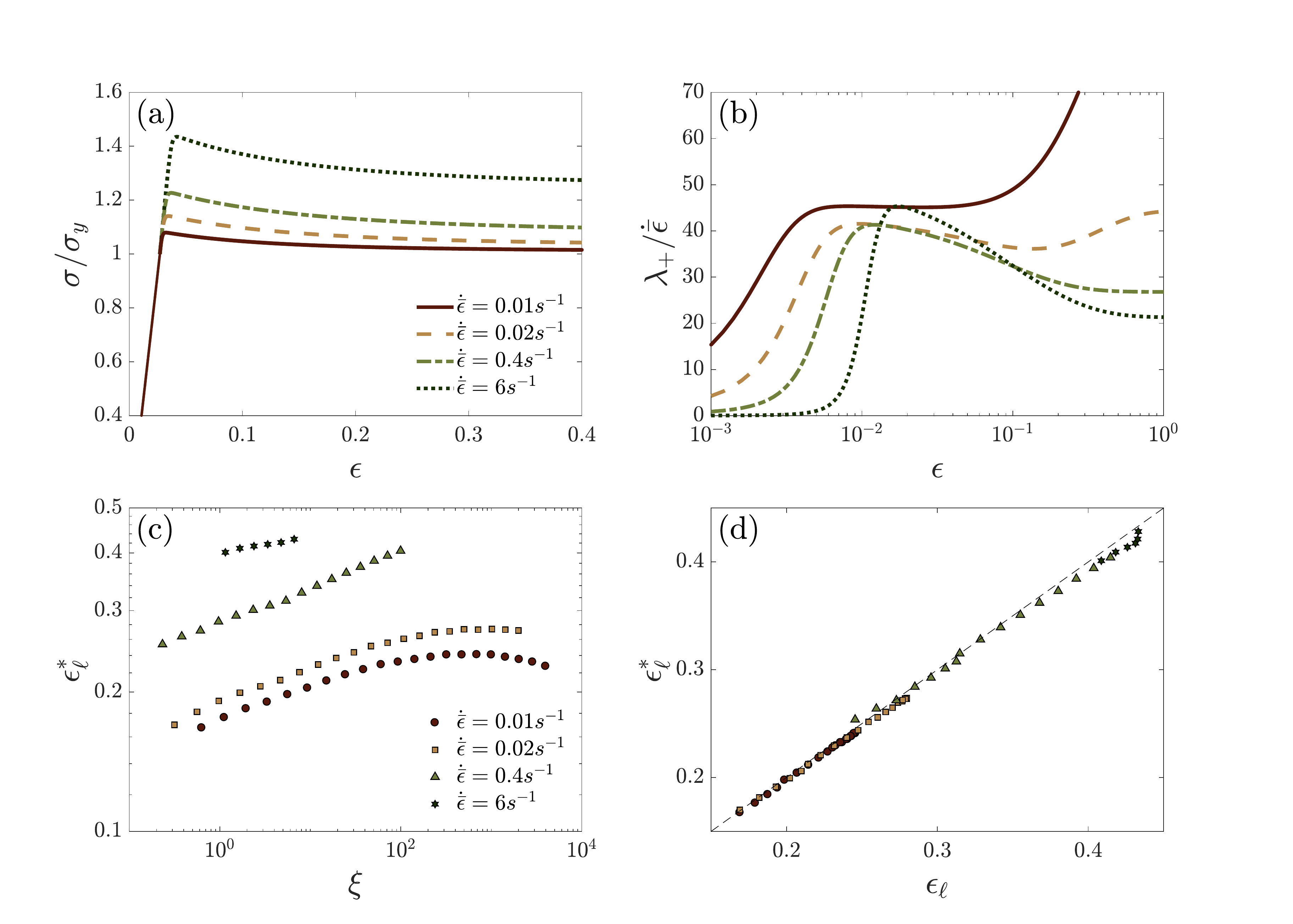}
\caption{(a) The normalized stress $\sigma/\sigma_y$ vs.~uniaxial strain $\epsilon$ corresponding to the spatially-homogeneous solution of the STZ model (see text for details) for $4$ imposed strain-rates, spanning nearly $3$ orders in magnitude (see legend). (b) The eigenvalue $\lambda_+$ of Eq.~\eqref{eq:lambda_+_generic} for the imposed strain-rates specified in panel (a). (c) The localization strain $\epsilon_{\ell}^*$ as obtained from full numerical solutions as a function of $\xi$, which quantifies the competition between the external loading rate and the initial plastic relaxation controlled by the glass structure/age (see text for details). (d) The numerically-obtained localization strain $\epsilon_{\ell}^*$ vs.~the predicted localization strain $\epsilon_{\ell}$ obtained from solutions of Eq.~\eqref{eq:strong_localization} with $\kappa\!=\!0.15$.}
\label{fig:4_panel_STZ}
\end{figure*}

We first consider the spatially-homogeneous solution of Eqs.~\eqref{eq:STZ_plastic},~\eqref{eq:STZ_IV} and~\eqref{eq:generic_1D_3}, where the latter reads $\dot\sigma/G\=\dot{\bar{\epsilon}}-\dot{\epsilon}^{pl}(\sigma,T,\chi)$. All calculations start at the onset of plasticity, $\sigma(t\=0)\=\sigma_y$, and the strain $\epsilon$ is measured relative to the elastic strain at this point. The stress-strain curves for various applied strain-rates $\dot{\bar{\epsilon}}$, spanning nearly $3$ orders in magnitude, are shown in Fig.~\ref{fig:4_panel_STZ}a. The calculations were performed with the following set of parameters: $G=37 \ \text{GPa}$ (shear modulus), $\tau_0\!=\!10^{-13} \ \text{s}$, $\sigma_y\!=\!0.85 \ \text{GPa}$, $e_z\simeq1.8 \ \text{eV}$, $\Delta\simeq0.69 \ \text{eV}$, $T\!=\!400$ K, $\bar{\Omega}\!=\!90 \ \text{\normalfont\AA}^3$, $c_0\!=\!0.4$, $\chi_{\infty}\!=\!900$ K, obtained in~\cite{Rycroft2015} for Vitreloy 1 (Vit1), the first commercial and widely-used bulk metallic glass. We observe that the stress typically exhibits a maximum that is followed by relaxation to steady-state, where the magnitude of the stress overshoot increases with increasing strain-rate. The stress peak is of elastic origin, i.e.~upon yielding the plastic strain-rate $\dot{\epsilon}^{pl}$ cannot keep up with the applied strain-rate $\dot{\bar{\epsilon}}$ and the latter is accommodated mainly by the elastic strain-rate, and the subsequent relaxation is of plastic origin as $\dot{\epsilon}^{pl}$ increases with increasing plastic deformation until reaching steady-state $\dot{\epsilon}^{pl}\!=\!\dot{\bar{\epsilon}}$.

A stress peak followed by stress relaxation is characteristic of amorphous/glassy materials which are strain-softening. As discussed in Sect.~\ref{sec:largest_eigenvalue}, for such materials we expect $\partial_{{\cal I}}\dot{{\cal I}}\!>\!0$, which implies that the largest eigenvalue in Eq.~\eqref{eq:lambda_+_generic} might be {\em positive} already at the early stages of the elasto-viscoplastic deformation process. This is directly demonstrated in Fig.~\ref{fig:4_panel_STZ}b, which also shows that the largest eigenvalue undergoes a significant evolution with strain, hence the onset conditions (i.e.~when $\lambda_+$ first becomes positive) cannot reasonably predict the strong localization process.

The results presented in Figs.~\ref{fig:4_panel_STZ}a-b can be used to test our criterion for strong localization in Eq.~\eqref{eq:strong_localization} against fully nonlinear numerical solutions. To that aim, we numerically solved Eqs.~\eqref{eq:generic_1D}, together with the constitutive law given by Eqs.~\eqref{eq:STZ_plastic}-\eqref{eq:STZ_IV} (see \hyperref[appendix]{Appendix} for details). Such solutions have already been presented in Fig.~\ref{fig:figure2}, where a sharp drop in the minimal cross-sectional area $a_{min}$ has been observed to occur at a strain $\epsilon_\ell^*$ (defined as the strain in which the slope is largest). We are interested in the dependence of the strong localization strain $\epsilon_\ell^*$ on the history of the glass or its initial age, as quantified by $\chi_0$, and on the imposed strain-rate $\dot{\bar{\epsilon}}$. Previous work~\cite{Vasoya2016} has shown that transient (i.e.~far from steady-state) elasto-viscoplastic deformation, which is of interest here, crucially depends on a competition between a timescale characterizing the external driving rate, $\tau^{ext}\!\equiv\!1/\dot{\bar{\epsilon}}$, and the {\em initial} plastic relaxation timescale, quantified by $\tau^{pl}\!\equiv\!\tau_0 \exp(-e_Z/k_B \chi_0)$ (cf.~Eq.~\eqref{eq:STZ_plastic}). Consequently, the dimensionless ratio $\xi\!\equiv\!\tau^{ext}/\tau^{pl}$ is expected to properly characterize transient elasto-viscoplastic deformation in this constitutive model.

In Fig.~\ref{fig:4_panel_STZ}c we plot $\epsilon_\ell^*$ as a function of $\xi$ for various applied strain-rates $\dot{\bar{\epsilon}}$, where $\xi$ is varied by up to $4$ orders of magnitude by varying $\chi_0$ for each $\dot{\bar{\epsilon}}$, and $\dot{\bar{\epsilon}}$ spans nearly $3$ orders of magnitude. It is observed that as the applied strain-rate $\dot{\bar{\epsilon}}$ is increased, strong localization occurs at larger strains $\epsilon_\ell^*$ for a fixed $\chi_0$ (i.e.~fixed initial structure and hence density of STZs), as observed experimentally for athermal amorphous systems such as bubble rafts~\cite{Kuo2013}, bulk metallic glasses~\cite{Vormelker2008}, and in previous computational studies~\cite{Eastgate2003,Eastgate2005}. In addition, as $\chi_0$ is increased for a fixed applied strain-rate $\dot{\bar{\epsilon}}$ --- i.e.~the initial viscoplastic response is stronger --- the system can sustain a larger strain $\epsilon_\ell^*$ prior to strong localization. This trend in fact depends on the strain-softening degree which is determined by the proximity of $\chi_0$ to the limiting value $\chi_\infty$ (cf.~Eq.~\eqref{eq:STZ_IV}); indeed, for the material parameters used here, when $\chi_0$ is sufficiently close to $\chi_\infty$, $\epsilon_\ell^*$ starts to decrease with increasing $\chi_0$ for the lowest applied strain-rate $\dot{\bar{\epsilon}}$, see lowest curve in Fig.~\ref{fig:4_panel_STZ}c.

The major question we are now ready to address is whether the strong localization strain $\epsilon_\ell^*$ measured in fully nonlinear extensional deformation solutions over a wide range of physical conditions, as presented in Fig.~\ref{fig:4_panel_STZ}c, is quantitatively predicted by the theoretical prediction in Eq.~\eqref{eq:strong_localization}. The latter is solved for $t_\ell$ based on the spatially-homogeneous solution used in Figs.~\ref{fig:4_panel_STZ}a-b, leading to $\epsilon_\ell\=\epsilon(t_\ell)$ once $\kappa\!\sim\!{\cal O}(10^{-1})$ is specified. In Fig.~\ref{fig:4_panel_STZ}d we plot the numerically measured $\epsilon_\ell^*$ against the predicted strong localization strain $\epsilon_\ell$ using $\kappa\=0.15$. It is observed that the theory quantitatively predicts the strong localization strain over a broad range of parameters using a single number $\kappa$, lending significant support to the proposed theoretical framework. In the next section we perform a similar analysis for a very different constitutive relation describing crystalline/polycrystalline materials.

\section{Application II: Crystal plasticity and the Kocks-Mecking model}
\label{sec:KM}

Encouraged by the success of the main theoretical result in Eq.~\eqref{eq:strong_localization} to quantitatively predict the strong localization strain in a constitutive model of amorphous/glassy materials, we set out to further test the degree of generality of the prediction. To that aim, we repeat the analysis of the previous section for a very different constitutive model. In particular, we consider the Kocks-Mecking model which has been rather widely used to describe plastic deformation in crystalline/polycrystalline materials~\cite{Mecking1981,Estrin1984,Follansbee1988,Kocks2003}. We choose to study this phenomenological model both because of its impact on the metallurgical literature and because it is one of the few models which explicitly invoke an internal-state field, an essential element in our theoretical framework.
\begin{figure*}
\centering
\includegraphics[width=0.88\textwidth]{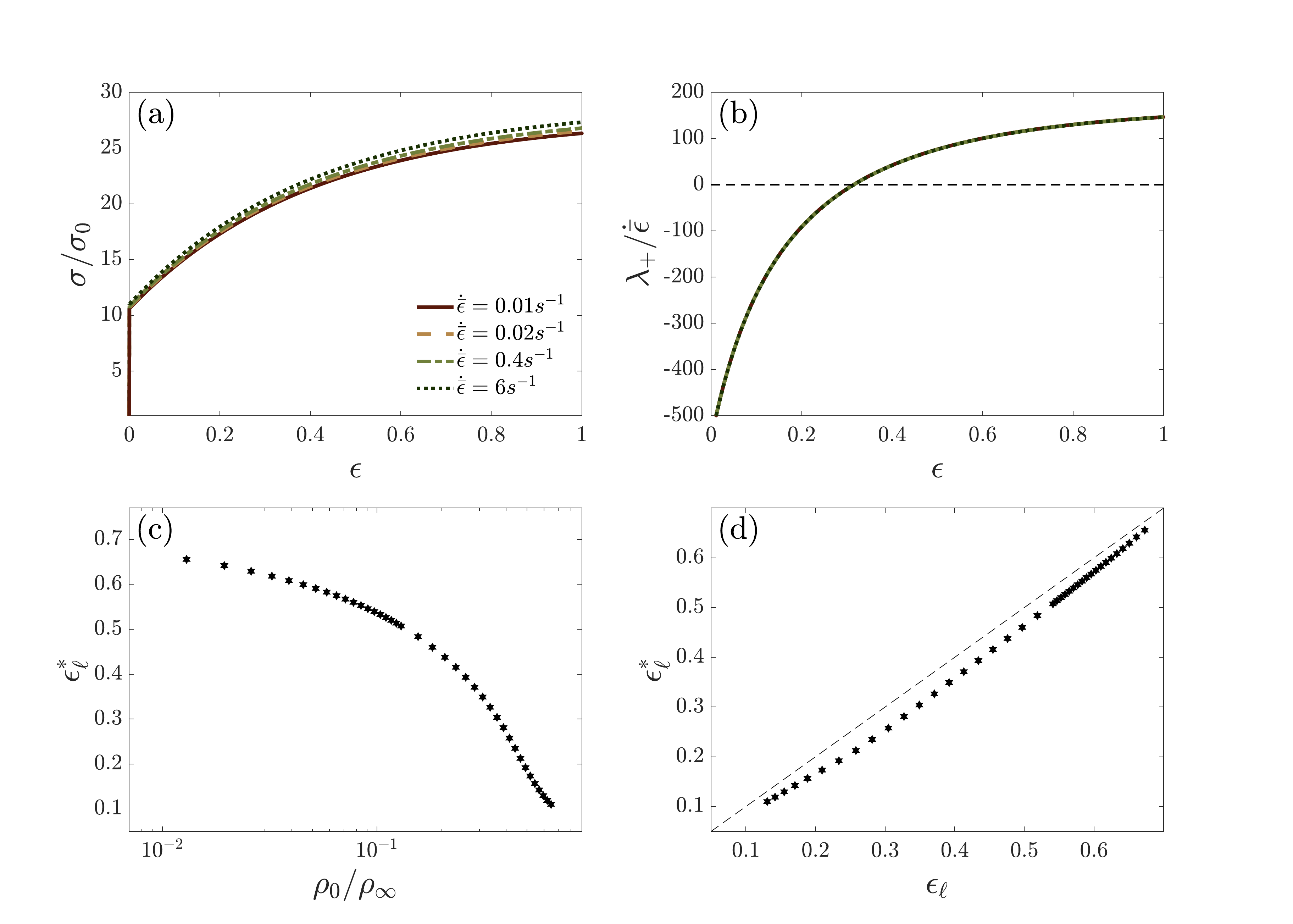}
\caption{(a) The normalized stress $\sigma/\sigma_0$ vs.~uniaxial strain $\epsilon$ corresponding to the spatially-homogeneous solution of the Kocks-Mecking model (see text for details) for $4$ imposed strain-rates, spanning nearly $3$ orders in magnitude (see legend). (b) The eigenvalue $\lambda_+$ of Eq.~\eqref{eq:lambda_+_generic} for the imposed strain-rates specified in panel (a). Note that on the scale used, the $4$ different curves appear to overlap. (c) The localization strain $\epsilon_{\ell}^*$ as obtained from full numerical solutions as a function of $\rho_0/\rho_{\infty}$ for a single strain-rate, $\dot{\bar{\epsilon}}=1 \ \text{s}^{-1}$ (results for the other strain-rates are not shown as they are practically indistinguishable). (d) The numerically-obtained localization strain $\epsilon_{\ell}^*$ vs.~the predicted localization strain $\epsilon_{\ell}$ obtained from a solution of Eq.~\eqref{eq:strong_localization} with $\kappa\!=\!0.1$.}
\label{fig:4_panel_KM}
\end{figure*}

The Kocks-Mecking constitutive model takes the form
\begin{subequations}
\label{eq:KM_plastic}
\begin{eqnarray}
  \dot{\epsilon}^{pl}(\sigma,\rho) &=& \tau_0^{-1} \left(\frac{\sigma - \sigma_0}{\sigma_T(\rho)}\right)^m\ ,\label{eq:KM_plasticA}\\
   \sigma_T(\rho) &=& \alpha_0\,M\,b\,G\,\sqrt{\rho} \ ,\label{eq:KM_plasticB}
\end{eqnarray}
\end{subequations}
which is somewhat analogous to Eqs.~\eqref{eq:STZ_plastic} of the STZ model of amorphous plasticity, yet it is very different in terms of the underlying physics. As the dominant microscopic mechanism of plastic deformation in crystalline materials is dislocation slip, the fundamental internal-state field in Eqs.~\eqref{eq:KM_plastic} is the areal density of dislocations $\rho$ (not to be confused with the mass density used earlier). It affects the plastic strain-rate in Eq.~\eqref{eq:KM_plasticA} through the Taylor-like stress $\sigma_T(\rho)$ in Eq.~\eqref{eq:KM_plasticB}, which quantifies the increase in the resistance to plastic deformation with increasing dislocation density $\rho$~\cite{Taylor1934a,Taylor1934b}. The physical idea is that dislocations remain locked-in the material and hamper the subsequent motion of other dislocations, which is believed to be at the heart of the important phenomenon of strain-hardening~\cite{Kocks2003}. $\sigma_T(\rho)$ in Eq.~\eqref{eq:KM_plasticB}, which scales with $\sqrt{\rho}$, is proportional to a constant $\alpha_0$ that depends on dislocation-dislocation interactions, the Taylor factor $M$, the magnitude of the Burger's vector $b$ and the shear modulus $G$ (these parameters depend on the temperature $T$, but we do not include this dependence explicitly). The plastic strain-rate $\dot{\epsilon}^{pl}(\sigma,\rho)$ in Eq.~\eqref{eq:KM_plasticA} is inversely proportional to a microscopic time $\tau_0$ and is driven by the applied stress $\sigma$ shifted by a reference flow-stress $\sigma_0$. Finally, the rate-sensitive relation between $\dot{\epsilon}^{pl}(\sigma,\rho)$ and the dimensionless stress $(\sigma-\sigma_0)/\sigma_T(\rho)$ is phenomenologically captured by the exponent $m$, consistently with the notation in Sect.~\ref{sec:largest_eigenvalue}.

The density of dislocations $\rho$ is a dynamical quantity that evolves with plastic deformation~\cite{Kocks2003}. Hence, to fully define the constitutive model, we need a dynamical $\rho$-evolution equation analogous to Eq.~\eqref{eq:STZ_IV}.
In the framework of the Kocks-Mecking model, it takes the form
\begin{equation}
\label{eq:KM_IV}
  \dot{\rho} = \dot{\epsilon}^{pl}\left(k_1 \sqrt{\rho} -k_2\,\rho\right) = k_2 \sqrt{\rho}\,\dot{\epsilon}^{pl} (\sqrt{\rho_\infty} -\sqrt{\rho})  \ ,
\end{equation}
where $\rho_{\infty}\!\equiv\!(k_1/k_2)^2$. Here $k_1 \sqrt{\rho}$ corresponds to the storage of dislocations and $k_2\,\rho$ corresponds to dynamic recovery, assuming the initial value of $\rho$ satisfies  $\rho_0\!<\!\rho_{\infty}$. Similarly to Eq.~\eqref{eq:STZ_IV}, the evolution of the material's internal structure is driven by plastic deformation, i.e.~it is proportional to $\dot{\epsilon}^{pl}$ in agreement with Eq.~\eqref{eq:internal_variable} and hence it also vanishes when $\dot{\epsilon}^{pl}\=0$. Note, though, that unlike Eq.~\eqref{eq:STZ_IV}, $\dot\rho$ is proportional to $\dot{\epsilon}^{pl}$, not to the plastic dissipation power $\sigma\dot{\epsilon}^{pl}$, which raises some questions about the proper generalization of Eq.~\eqref{eq:KM_IV} to non-unidirectional and to fully tensorial viscoplastic flows, as discussed in~\cite{Langer2010}.

Equations~\eqref{eq:KM_plastic}-\eqref{eq:KM_IV} have been recently analyzed in a similar context in~\cite{Yasnikov2014,Yasnikov2017}. The analysis in~\cite{Yasnikov2014,Yasnikov2017}, however, exclusively focussed on the onset conditions for necking. Indeed, applying our {\em general} expression for the largest eigenvalue in Eq.~\eqref{eq:lambda_+_generic} to Eqs.~\eqref{eq:KM_plastic}-\eqref{eq:KM_IV}, we recover the onset criterion derived in~\cite{Yasnikov2017} for the Kocks-Mecking model (cf.~Table 2 in the main text of~\cite{Yasnikov2017} and Eq.~(A11) in its Appendix/Supplementary data). Yet, our analysis goes well beyond that of~\cite{Yasnikov2014,Yasnikov2017}, not only in providing a general onset criterion that is not specific to a particular constitutive relation, i.e.~Eq.~\eqref{eq:lambda_+_generic}, but more importantly, in predicting the emergence of strong localization, which is not discussed at all in~\cite{Yasnikov2014,Yasnikov2017}.

To test our prediction for the emergence of strong localization in the Kocks-Mecking model, we repeated the analysis of the previous section with Eqs.~\eqref{eq:KM_plastic}-\eqref{eq:KM_IV} instead of Eqs.~\eqref{eq:STZ_plastic}-\eqref{eq:STZ_IV}. All calculations start at the onset of plasticity (see \hyperref[appendix]{Appendix} for details), and the strain $\epsilon$ is measured relative to the elastic strain at this point. The calculations were performed with the following set of parameter: $G\!=\!48\ \text{GPa}$, $\tau_0\!=\!10^{-6} \ \text{s}$, $m\!=\!167$, $\sigma_0\!=\!13.4 \ \text{MPa}$, $\alpha_0 M b\!=\!3\times 10^{-10} \ \text{m}$, $k_1\!=\!139\times 10^{6} \ \text{m}^{-1}$, $k_2\!=\!5$, characteristic of copper processed by equal channel angular extrusion~\cite{Yasnikov2017, DallaTorre2004}. The spatially-homogeneous stress-strain curves for various applied strain-rates $\dot{\bar{\epsilon}}$, shown in Fig.~\ref{fig:4_panel_KM}a, exhibit a strain-hardening behavior characteristic of crystalline materials. That is, the stress in these curves increases monotonically with strain, a behavior that is qualitatively different from the strain-softening behavior typically exhibited by amorphous materials, cf.~Fig.~\ref{fig:4_panel_STZ}a. Note that as the strain-hardening ``rate'' $d\sigma/d\epsilon$ in the plastic regime is significantly smaller than the elastic modulus $G$, the elastic branch in the stress-strain curves appears to be nearly vertical in Fig.~\ref{fig:4_panel_KM}a. Also note that the exhibited strain-rate sensitivity is quite small.

As discussed in Sect.~\ref{sec:largest_eigenvalue}, strain-hardening materials typically feature $\partial_{{\cal I}}\dot{{\cal I}}\!<\!0$ (here ${\cal I}$ is the dislocation density $\rho$), which implies that the expression in Eq.~\eqref{eq:lambda_+_generic} may be {\em negative} during the elasto-viscoplastic deformation process, in which case the largest eigenvalue is taken to be $\lambda_+\=0$. As shown in the examples in Fig.~\ref{fig:4_panel_KM}b, $\lambda_+$ corresponding to Eq.~\eqref{eq:lambda_+_generic} is indeed negative in the early stages of the extensional deformation; consequently, we take the integrand of Eq.~\eqref{eq:strong_localization} to vanish in this range of strains. At larger strains, the largest eigenvalue becomes positive and sizable, as shown in Fig.~\ref{fig:4_panel_KM}b, and a necking instability develops (in this range the integrand of Eq.~\eqref{eq:strong_localization} is used as is).

The strong localization strain $\epsilon_l^*$ is shown in Fig.~\ref{fig:4_panel_KM}c as a function of the initial density of dislocations $\rho_0$ for a single strain-rate, $\dot{\bar{\epsilon}}\=1 \ \text{s}^{-1}$ (results for the other strain-rates are not shown as they are practically indistinguishable). Interestingly, and somewhat nonintuitively, $\epsilon_l^*$ is a decreasing function of $\rho_0$ in this model, a result that should be tested against experimental data (we are not aware of existing data that quantify the variation of neck evolution with the initial density of dislocations). In Fig.~\ref{fig:4_panel_KM}d we plot the numerically measured $\epsilon_\ell^*$ against the predicted strong localization strain $\epsilon_\ell$ using $\kappa\=0.1$. It is observed that yet again the theory reasonably well predicts the strong localization strain using a single number $\kappa$. These results, together with the results of the previous section, seriously support the theoretical framework developed in this paper and open the way to better understanding necking instabilities in elasto-viscoplastic materials.

\section{Concluding remarks}
\label{sec:summary}

We have presented a rather general analysis of necking instabilities in rate-dependent elasto-viscoplastic materials in the long-wavelength approximation, within a generic constitutive framework that accounts for the structural evolution of the material during plastic deformation through an internal-state field. Developing an approximated WKB-like time-dependent linear stability analysis, we derived in Eq.~\eqref{eq:lambda_+_generic} an analytical expression for the largest time-dependent eigenvalue in the stability problem, which reveals the various destabilizing and stabilizing physical processes involved in necking instabilities. The onset of necking criterion, associated with the strain/time in which the largest eigenvalue of Eq.~\eqref{eq:lambda_+_generic} becomes positive, is discussed in relation to onset criteria available in the literature.

Our theoretical framework, however, allows to go significantly beyond the onset condition in order to derive a criterion for the emergence of strong localization, when the cross-sectional area reduction associated with necking becomes macroscopically appreciable. The analytical criterion, presented in Eq.~\eqref{eq:strong_localization}, is tested against fully nonlinear numerical solutions of two widely different elasto-viscoplastic constitutive relations involving an internal-state field, one for strain-softening amorphous/glassy materials (using the Shear-Transformation-Zone model) and the other for strain-hardening crystalline/polycrystalline materials (using the Kocks-Mecking model). Both analyses demonstrate that the criterion in Eq.~\eqref{eq:strong_localization} quantitatively predicts the onset of strong localization, exhibiting weak (logarithmic) dependence on the typical (relative) magnitude of initial perturbations $\zeta$ and a dimensionless parameter $\kappa$ of ${\cal O}(10^{-1})$. Consequently, we believe that our theoretical results can be used to predict necking instabilities in a broad range of materials and constitutive relations.

Our results open up the way for several future research directions, some of which are mentioned here. First, our general results should be applied to more sophisticated constitutive relations, for example those that are capable of quantitatively describing the elsato-viscoplastic flows of glasses near their glass temperature. This is highly relevant, for example, for many bulk metallic glass processing methods, such blow molding or stretch rolling, where necking instabilities play a critical role. Second, our predictions should be tested against experimental data for various materials, where more refined measurements of the evolution of the material microstructure during plastic deformation should be performed in order to guide the choice of the relevant of internal-state fields and their dynamics. Finally, the range of validity and limitations of the long-wavelength (lubrication or slender-bar) approximation should be elucidated. To this aim, one needs to consider two-dimensional and probably also three-dimensional extensions of the present analysis, which most likely entail large-scale numerical calculations involving advanced computational techniques. Such calculations should also resolve the role of boundary effects, e.g.~those associated with clamped boundary conditions. Obviously, these issues are not only related to the range of validity of the presented theory, but also to quantitative comparisons to experiments.

\acknowledgements
We acknowledge support from the Richard F.~Goodman Yale/Weizmann Exchange Program, the William Z.~and Eda Bess
Novick Young Scientist Fund and the Harold Perlman Family. We are grateful to Y.~Bar-Sinai for his valuable help in developing the numerical tools, to E.A.~Brener for fruitful discussions and a critical reading of the manuscript, and to C.H.~Rycroft and J.~Schroers for their comments on the final version of the manuscript.

\appendix*
\section{Fully nonlinear numerical solutions}
\label{appendix}

During extensional deformation, the material changes its shape, mathematically implying a time-dependent, evolving domain. In order to obtain fully nonlinear solution of the problem at hand, we performed a coordinate transformation from the Eulerian (spatial) frame of reference $x$ to the Lagrangian (material) frame of reference $X$. In the latter frame of reference, the domain is fixed (i.e.~time-independent), but the equations contain additional nonlinearities. In particular, Eqs.~\eqref{eq:generic_1D} take the form
\begin{subequations}
\label{eq:generic_1D_Lag}
    \begin{align}
   0 &= J^{-1}\partial_X\left(A\Sigma\right)\equiv J^{-1}\partial_XF\ ,\label{eq:generic_1D_Lag_1}\\
  \partial_t A &= -A \frac{\partial_t J}{J} \ ,\label{eq:generic_1D_Lag_2} \\
  \partial_t F &= G A \left(\frac{\partial_t J}{J}-\dot{E}^{pl}\right)+F \,\frac{\partial_t A}{A} \ ,\label{eq:generic_1D_Lag_3} \\
  \partial_t {\cal\hat{I}} &=\dot{E}^{pl}\,\mathcal{G}\left(\Sigma,{\cal\hat{I}}\right) \ .\label{eq:generic_1D_Lag_4}
  \end{align}
\end{subequations}

In Eqs.~\eqref{eq:generic_1D_Lag}, $J\!\left(X,t\right)\!\equiv\!\partial x/\partial X$ is the Jacobian of the transformation (the scalar counterpart of the deformation gradient tensor), and $A\!\left(X,t\right),\Sigma\!\left(X,t\right)$, ${\cal\hat{I}}\!\left(X,t\right)$ and $V\!\left(X,t\right)$ are the Lagrangian counterparts of the Eulerian fields $a\!\left(x,t\right),\sigma\!\left(x,t\right)$, ${\cal I}\!\left(x,t\right)$ and $v\!\left(x,t\right)$ in Eqs.~\eqref{eq:generic_1D}. While the latter are defined over a time-dependent spatial domain $x\!\in\!\left[-\tfrac{L\left(t\right)}{2},\tfrac{L\left(t\right)}{2}\right]$, the former are defined over a fixed (time-independent) spatial domain $X\!\in\!\left[-\tfrac{L_0}{2},\tfrac{L_0}{2}\right]$. $\dot{E}^{pl}(X,t)$ is the Lagrangian counterpart of the Eulerian plastic strain-rate $\dot{\epsilon}^{pl}(x,t)$, and ${\cal G}(\Sigma(X,t),{\cal\hat{I}}(X,t))$ is the Lagrangian counterpart of the Eulerian function $g(\sigma(x,t), {\cal I}(x,t))$ in Eq.~\eqref{eq:generic_1D_4}. The convective derivative $\partial_t+v\partial_x$ in the Eulerian formulation is simply replaced by a partial time derivative $\partial_t$ in the Lagrangian formulation. Finally, note that $\partial_X V\!=\!\partial_t J$, which implies $\partial_x v\!\rightarrow\!\partial_t J/J$.

Equation~\eqref{eq:generic_1D_Lag_1} is simply solved by a space-independent force, $F(t)\!=\!A(X,t)\,\Sigma(X,t)$. The remaining equations, Eqs.~\eqref{eq:generic_1D_Lag_2}-\eqref{eq:generic_1D_Lag_4}, are solved numerically. The fields $A(X,t)$, $J(X,t)$ and ${\cal\hat{I}}(X,t)$ are discretized in space using the Method of Lines~\cite{schiesser2012}. At each point in time, each field is represented by $N$ discrete values defined at $N$ spatial points and $F$ is characterized by a single space-independent value. The total $3N+1$ degrees of freedom is constrained by only $3N$ equations, where each of Eqs.~\eqref{eq:generic_1D_Lag_2}-\eqref{eq:generic_1D_Lag_4} contributes $N$ equations. The additional equation emerges from the boundary condition. In particular, as we impose the velocity at the bar's edges, $v(x\!=\!L(t)/2,t)\!=\!-v(x\!=\!-L(t)/2,t)$, in the Eulerian formulation we have
\begin{equation}
\label{eq:boundary_cond_Eulerian}
  \left\langle \partial_x v\right\rangle =\frac{1}{L(t)}\!\int_{-\frac{L(t)}{2}}^{\frac{L(t)}{2}}\!\partial_x v(x,t) \,dx = \frac{2v(x\!=\!L(t)/2,t)}{L(t)}= \dot{\bar{\epsilon}} \ .
\end{equation}
Transforming Eq.~\eqref{eq:boundary_cond_Eulerian} to the Lagrangian frame of reference, we obtain
\begin{equation}
\label{eq:boundary_cond_Lagrangian}
  \frac{1}{L_0}\int_{-\frac{L_0}{2}}^{\frac{L_0}{2}} \partial_t J(X,t) \,dX =  \dot{\bar{\epsilon}}\,e^{\dot{\bar{\epsilon}}t} \ .
\end{equation}
This integral relation, which adds spatial coupling to the ordinary time differential Eqs.~\eqref{eq:generic_1D_Lag_2}-\eqref{eq:generic_1D_Lag_4}, is the extra equation we needed. Equation \eqref{eq:boundary_cond_Lagrangian} was discretized using a simple sum in the numerical implementation.

The spatially discretized system of $3N+1$ equations was integrated in time using
a standard differential-algebraic equation solver in Mathematica~\cite{mathematica} with the following initial conditions
\begin{equation}
\label{eq:init_cond_Lagrangian}
    \begin{aligned}
  A \left(X,0\right) &= 1\ , \\
  F \left(0\right) &= A(X,0)\,\Sigma(X,0) \ ,  \\
  {\cal\hat{I}} \left(X,0\right) &= {\cal\hat{I}}_0\left(1+\zeta\cos\left(2\pi X/L_0\right)\right)\ ,
  \end{aligned}
\end{equation}
where $\Sigma(X,0)$ should be specified (see below), $\zeta$ is the relative amplitude of the perturbation and ${\cal\hat{I}}_0$ is a constant initial background level of the internal-state field. The time $t\=0$ is set to the onset of plastic deformation, which is also used to define the Lagrangian frame of reference, $J \left(X,0\right)\= 1$ (i.e.~the small initial elastic deformation is neglected). Consequently, the initial force $F(t\=0)$ is set by $\Sigma(X,0)\!=\!\sigma_y$ for the STZ model and by $\Sigma(X,0)$ that satisfies $\dot{\bar{\epsilon}}\!=\!\dot{E}^{pl}\left(\Sigma(X,0),\mathcal{R}_0\right)$, where $\mathcal{R}_0$ is the Lagrangian counterpart of initial dislocation density $\rho_0$, for the Kocks-Mecking model. In all presented results we used $N\=1000$ (higher values of $N$ were used to verify the appropriate numerical convergence associated with this choice).


%

\end{document}